\documentclass[reprint,floatfix,superscriptaddress,twocolumn,footinbib]{revtex4-2}

\usepackage{pifont} 

\usepackage[utf8]{inputenc}
\usepackage[a4paper]{geometry}
\usepackage[colorlinks,linktocpage,bookmarks=false]{hyperref} 
\usepackage[table,dvipsnames]{xcolor}
\usepackage{color}
\usepackage{multirow}
\usepackage{ulem}

\usepackage{bm}
\usepackage{comment}
\usepackage{enumitem}
\usepackage[version=4]{mhchem}
\usepackage{setspace}
\usepackage[caption=false]{subfig}
\usepackage[T1]{fontenc}

\usepackage{amsmath,amssymb}
\usepackage{amsfonts, relsize}
\usepackage{mathtools}
\usepackage{physics}
\usepackage{bbm}

\newcommand\myshade{85}
\definecolor{myrulecolor}{RGB}{150,20,0}
\colorlet{mylinkcolor}{violet}
\colorlet{mycitecolor}{YellowOrange}
\colorlet{myurlcolor}{Aquamarine}
\hypersetup{
	linkcolor  = myurlcolor!\myshade!black,
	citecolor  = mycitecolor!\myshade!black,
	urlcolor   =  myrulecolor!\myshade!black, 
}

\newcommand{\rucl}{\ce{\alpha-RuCl3}}

\newcommand{\eps}{\varepsilon}

\newcommand{\beq}{\begin{equation}}
\newcommand{\eeq}{\end{equation}}
\renewcommand{\emph}{\textit}

\makeatletter
\newcommand*{\rom}[1]{\expandafter\@slowromancap\romannumeral #1@}
\makeatother

\begin{document}
\newcommand{\thistitle}{Magnons, Phonons, and Thermal Hall Effect in Candidate Kitaev Magnet $\alpha$-RuCl$_3$}

\title{\thistitle}

\author{Shuyi Li}
\affiliation{Department of Physics and Astronomy, Rice University, Houston, TX 77005, USA}

\author{Han Yan}
\affiliation{Department of Physics and Astronomy, Rice University, Houston, TX 77005, USA}

\affiliation{Smalley-Curl Institute, Rice University, Houston, TX 77005, USA}

\author{Andriy H. Nevidomskyy}
\email[Correspondence e-mail address: ]{nevidomskyy@rice.edu}
\affiliation{Department of Physics and Astronomy, Rice University, Houston, TX 77005, USA}

\date{\today}

\begin{abstract}
We study the nature of the debated thermal Hall effect in the candidate Kitaev material \rucl. Without assuming the existence of a gapped spin liquid, we show that a realistic minimal spin model
in the canted zigzag phase suffices, at the level of linear spin-wave theory,  to qualitatively explain  the observed temperature and magnetic field dependence of the non-quantized thermal Hall conductivity $\kappa_{xy}$, with its origin lying in the Berry curvature of the magnon bands. 
The magnitude of the effect is however too small  compared to the measurement 
by Czajka \textit{et al.} [\href{https://doi.org/10.1038/s41563-022-01397-w} {Nat. Mater. \textbf{22},  36–41 (2023)}], 
even after 
scanning a broad range of model parameters so as to maximize $\kappa_{xy}/T$. Recent experiments suggest that phonons play an important role, which we show couple to the spins, endowing phonons with chirality. The resulting intrinsic contribution, from both magnons and phonons, is however still insufficient to explain the observed magnitude of the Hall signal. 
After careful analysis of the extrinsic phonon mechanisms, we use the recent experimental data on thermal transport in $\alpha$-RuCl$_3$ by Lefran\c{c}ois \textit{et al.} [\href{https://doi.org/10.1103/PhysRevX.12.021025}{Phys. Rev. X \textbf{12}, 021025 (2022)}] to determine the phenomenological ratio of the extrinsic and intrinsic contributions $\eta\equiv \kappa_{xy}^{E}/\kappa_{xy}^{I}$. We find $\eta=1.2\pm 0.5$, which when combined with our computed intrinsic value, explains quantitavely both the  magnitude and detailed temperature dependence of the experimental thermal Hall effect in \rucl.
\end{abstract}


\maketitle

The proposal that a quasi-2D Mott insulator \rucl\  may provide a realization~\cite{Plumb2014} of Kitaev's celebrated honeycomb compass model~\cite{kitaev2006} has attracted much attention to this material. While \rucl\  orders antiferromagnetically below $T_N= 7$~K~\cite{cao2016}, it was found that an in-plane magnetic field $h\equiv\mu_0H_\parallel \lesssim 10$~T is sufficient to suppress the magnetic order. While the nature of the resulting phase is still under intense debate, the observation of approximately quantized value of the thermal Hall conductivity $\kappa_{xy}/T$ in a narrow range of field ($6<h<9$~T)~\cite{Matsuda-quantized,Matsuda-quantized2} was attributed to the presence of the Majorana edge mode, predicted to exist in Kitaev's spin liquid subjected to an external magnetic field~\cite{kitaev2006,Nasu2017}. 
This interpretation has been recently challenged by an independent measurement of the thermal Hall effect~\cite{ONG}, in which the authors find a non-quantized, temperature-dependent $\kappa_{xy}$, which they attribute to a bosonic, rather than fermionic mechanism \cite{Mcclarty2018,CookmeyerPRB2018,Chern2021PRL,YBK,Li-Okamoto2022}.
Its nature  remains controversial, with one recent experimental study suggesting the possibility of quantized Hall effect in high fields $h>10$~T~\cite{Takagi-quantized2022}, while another attributing the origin of the thermal Hall effect to phonons~\cite{taillefer-phonons2022}.

In this Letter, we investigate the possibility of the bosonic origin of thermal conductivity in a widely accepted spin model of \rucl.  Since there is a considerable debate on the precise values of the model parameters describing \rucl, we perform a careful scan over a wide region in the parameter space to determine the largest possible values of $\kappa_{xy}/T$. We find that the bulk magnon excitations alone cannot explain the experimentally measured values of thermal conductivity in \rucl, even under the most favourable circumstances. Instead, we find that it is crucial to take the magneto-elastic coupling into consideration, whereby acoustic phonons hybridize with the magnon excitations, boosting the value of $\kappa_{xy}/T$. Even then, it turns out that in order to explain the experimental measurements, one must consider not only intrinsic but also extrinsic contributions to the thermal Hall effect, such as the skew-scattering of phonons/magnons off of impurities. We deduce the realistic value of this extrinsic contribution from a recent measurement on \rucl. When magnon, phonon and extrinsic contributions are taken into account, we are able to quantitatively reproduce the recent experimental data~\cite{ONG} on thermal conductivity in this material.

\textit{Model and Phases}.  $\alpha$-RuCl$_3$ 
emerged as a candidate material to study Kitaev physics on the honeycomb lattice because of its purported proximity to the spin-liquid state~\cite{Plumb2014}. In addition to the Kitaev's bond-dependent interactions stemming from the interplay of spin-orbit coupling and superexchange between Ru$^{3+}$ ions~\cite{Jackeli2009}, the importance of nearest-neighbor Heisenberg interactions $J_1$ and the off-diagonal exchanges, so-called $\Gamma$ and $\Gamma'$ terms~\cite{Rau2014} has been established.
Much theoretical work~\cite{Kim2015,Kim2016,Winter2016,Chaloupka2016,Yadav2016,Winter2017,Hou2017,Winter2018,CookmeyerPRB2018,Eichstaedt2019,Laurell2020,SashaC} has since focused on deducing the values of these parameters in \rucl, leading to the minimal effective spin-$\frac{1}{2}$ model of the form~\cite{Kim2016,Winter2016,Janssen-Vojta2017,Suzuki_models2018,Laurell2020,Chern2021PRL}

\vspace{-2mm}
\begin{align}
  H_{\text{m}}=&\sum_{\langle ij\rangle_1\in\alpha}[J_1\vec{S}_i\cdot\vec{S}_j+KS_i^{\alpha}S_j^{\alpha}+\Gamma(S_i^{\beta}S_j^{\gamma}+S_i^{\gamma}S_j^{\beta})\nonumber\\
&+\Gamma'(S_i^{\alpha}S_j^{\beta}+S_i^{\beta}S_j^{\alpha}+S_i^{\gamma}S_j^{\alpha}+S_i^{\alpha}S_j^{\gamma})]\nonumber\\
  &+\sum_{\langle ij\rangle_3}J_3\vec{S}_i\cdot\vec{S}_j-\sum_i \frac{g\mu_B}{\hbar}\vec{h}\cdot\vec{S}_i,\label{eq.model}
\end{align}
where the third-neighbour Heisenberg exchange $J_3$~\cite{Winter2017} was also added. 
The index $\alpha=(x, y, z)$ enumerates the three nearest bonds on the honeycomb lattice and also labels the bond-dependent spin couplings, with the remaining indices $\beta,\gamma$ taking values among the cyclic permutations of $(x,y,z)$ indices, for a given $\alpha$ (see SM).

Since the experiments are conducted under the applied magnetic field along the  $a$-axis, its effect is captured by the last term in Eq.~\eqref{eq.model} with the Land\'e g-factor $g=2.5$ \cite{KubotaPhysRevB2015,Yadav2016,Winter2018}. It is important to emphasize that the Kitaev axes $(x,y,z)$ are the so-called cubic axes~\cite{Winter2017} that do not coincide with the crystallographic ones. 
%
In particular, the magnetic field along the $a$-axis has nonzero  components along all three Cartesian $x, y, z$ axes, which in the pure Kitaev model is predicted to open a spectral gap proportional to the third power of the field~\cite{kitaev2006}.

\begin{figure}[!t]
    \centering
    \subfloat{\includegraphics[width=0.35\textwidth]{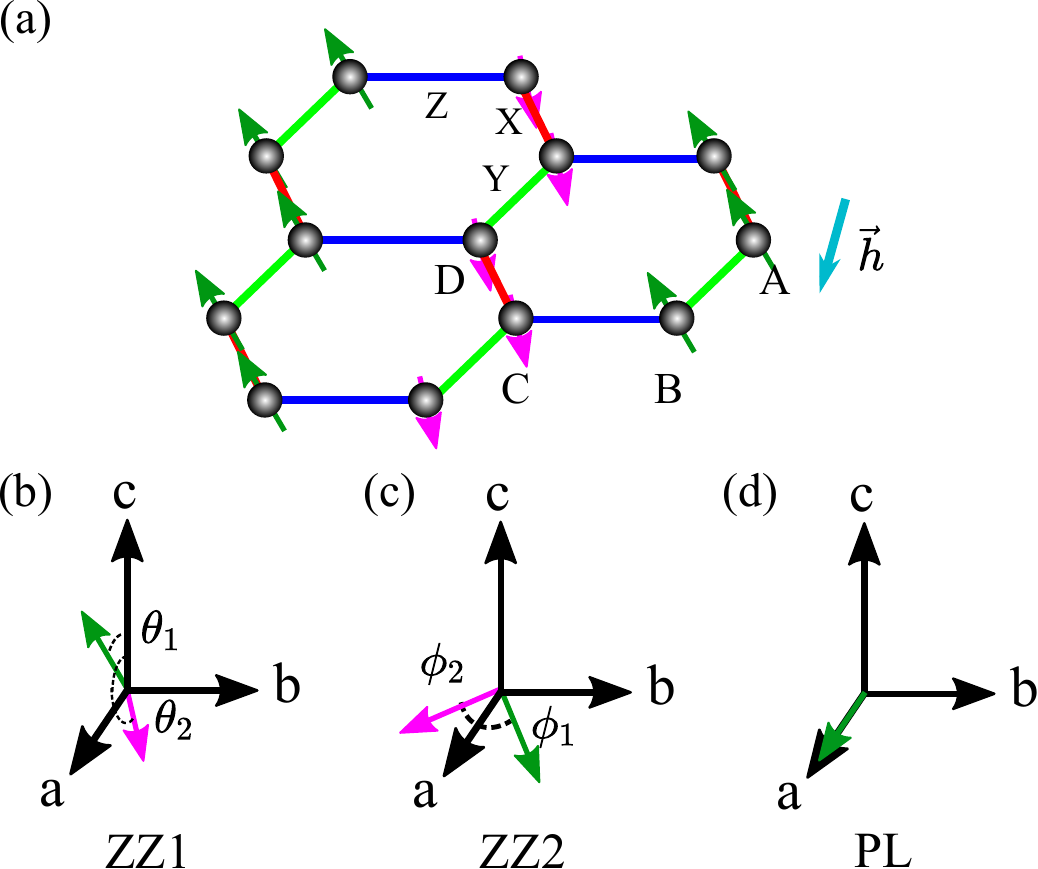}}\\
    \subfloat{\includegraphics[width=0.35\textwidth]{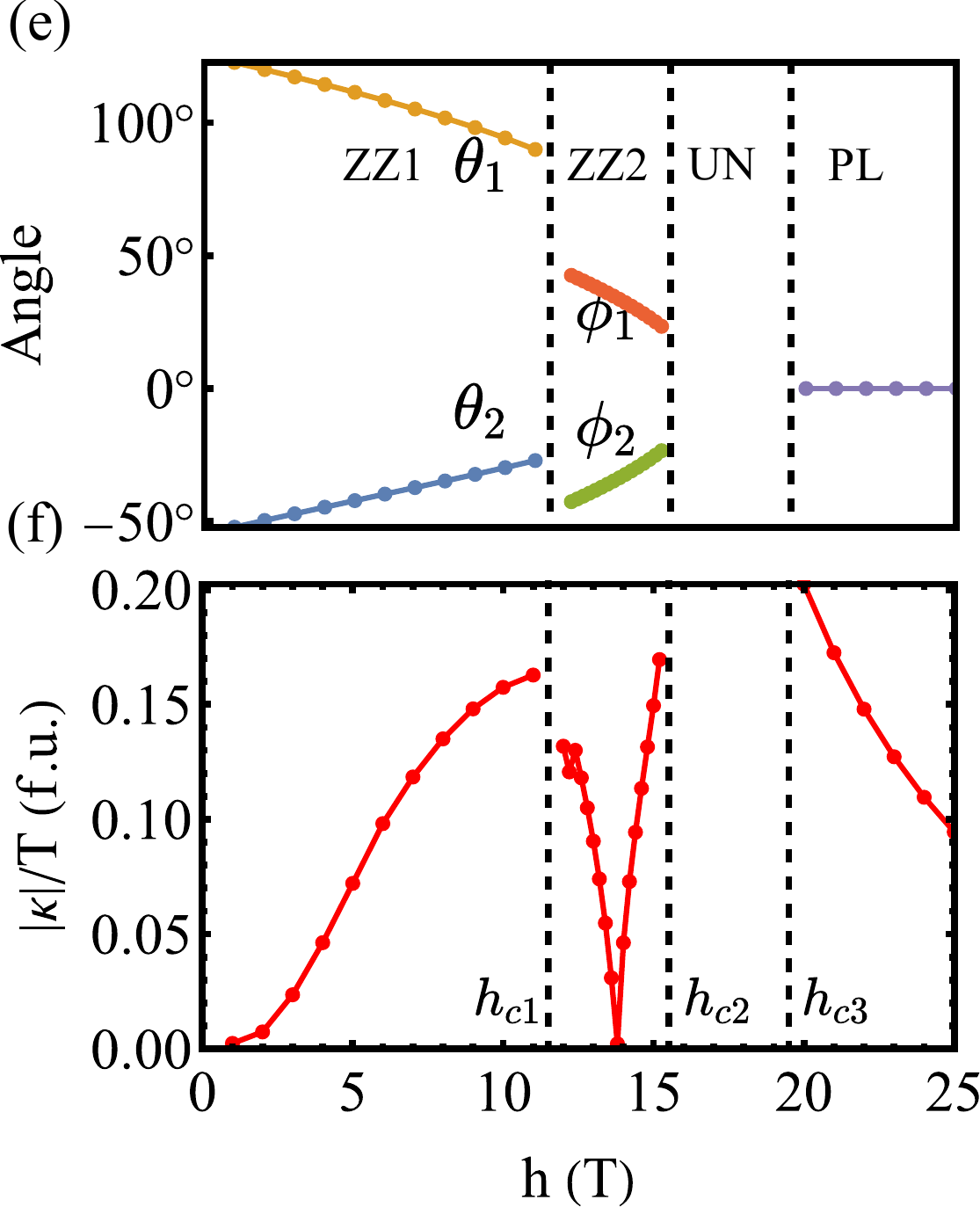}}
    \vspace{-2mm}
    \caption{
    (a) The lattice structure of \rucl with only Ru$^{3+}$ ions depicted for clarity and the nearest bonds $\alpha=\{X,Y,Z\}$ shown in red, blue and green colors. 
    (b-d): angles of the spins on sublattices A and B in the canted ZZ1 and ZZ2 phases, and in the fully polarized (PL) phase.
    (e) The angles of the spins relative to the $a$ axis shown as functions of increasing magnetic field, using the parameter set $(K,J_1,J_3,\Gamma,\Gamma')=(-7.2,0,0.8,0,-0.2)$ meV 
    (UN labels the unknown phase not captured by the zigzag ansatz). (f) Evolution of the thermal Hall conductivity $|\kappa^{2D}_{xy}|/T$ (in fermionic units $\pi k_B^2/6\hbar$) at $T=10$ K witihin the linear spin-wave theory.
    \vspace{-5mm}}
    \label{fig.kgmodel}
\end{figure}

In all the parameter sets proposed in previous works, obtained either from first-principles calculations or from phenomenological analysis (see e.g. Ref.~\cite{SashaC} for review), the leading coupling is believed to be the \textit{ferromagnetic} Kitaev term $K<0$~\cite{Sears_fm-Kitaev_exp2020}, with the off-diagonal term $\Gamma>0$ large and potentially comparable to $|K|$. 
In what follows, we assume the strength of the Kitaev interaction to be $K=-7.2$ meV, which is close to a recent \textit{ab initio} derived value of 80~K and is in the middle of the ``realistic parameter regime'' proposed in Ref.~\cite{SashaC}. The subleading Heisenberg interactions $J_1<0$ and $J_3>0$ are also necessary to explain the ordered zigzag phase of \rucl. The behaviour of this model  in the applied field is illustrated in Fig.~\ref{fig.kgmodel} for a representative parameter set $(K,J_1,J_3,\Gamma,\Gamma')=(-7.2,0,0.8,0,-0.2)$ meV. 
As the strength of the magnetic field (along the $a$-axis) increases, 
the spins tilt along the field direction, resulting in the canted zigzag phases ZZ1 and ZZ2 depicted schematically in Figs.~\ref{fig.kgmodel}(b,c) -- what distinguishes these two phases is the plane in which the spins of the two sublattices lie. 
At a sufficiently large field (whose value depends on the model parameters, and here $h_\text{sat}=19$~T), a fully polarized (PL) phase is reached. 
We use the standard linear spin wave theory (LSWT) (see Supplementary Materials (SM)) to compute the magnon spectrum of the model, and hence the thermal conductivity, given by the well known formula~\cite{MagnonHEth1,MagnonHEth2,MagnonHEth3,MagnonHEth5,Murakami2016}: 
\begin{equation}
    \kappa_{xy}^{3D}=-\frac{k_B^2T}{(2\pi)^3\hbar}\int_{BZ(3D)}d\vec{k}\sum_nc_2(f(\epsilon_{n\vec{k}}))\Omega_{xy}^n(\vec{k}), 
\label{eq.kxy}
\end{equation}
where $c_2(x)=(1+x)(\ln[(1+x)/x])^2-(\ln x)^2-2Li_2(-x)$, $Li_2(x)$ is the dilogarithm function,   $f(\epsilon)$ is the Bose-Einstein distribution, and the summation is over all the magnon bands. To compare with the prediction of the two-dimensional Kitaev QSL originating from the Majorana edge modes: $\kappa_{xy}^{2D}/T=\pi k_B^2/12\hbar$, we compute the same quantity in the unit of fermionic quantized value $\pi k_B^2/6\hbar$:
\begin{equation}
\label{eq.kxyfu}
\left.\frac{\kappa_{xy}^{2D}}{T}\right\vert_{\text{f.u.}}=\frac{\kappa_{xy}^{3D}d}{T}\times\frac{6\hbar}{\pi k_B^2}=\frac{3}{2\pi^3}\!\sum_{n,\vec{k}\in BZ(2D)}\!\!c_2(f(\epsilon_{n\vec{k}_0}))\phi_{n\vec{k}},
\end{equation}
where $d=5.72\text{ \r{A}}$ is the interlayer distance of $\alpha$-RuCl$_3$. 
In these fermionic units (f.u.), the quantized value reported in Ref.~\cite{Matsuda-quantized} would be $\kappa_{xy}^\text{Maj}=0.5$~f.u. 
We compute the integral by summing over the Berry flux $\phi_{n\vec{k}}$ in each small plaquette of the Brillouin zone (BZ) 
\cite{bcapprox1,bcapprox2}, weighted by the $c_2$ function (see SM for more details). The resulting (magnon only) thermal Hall conductivity $|\kappa^{2D}_{xy}|/T$ is plotted, at $T=10$~K, as a function of increasing magnetic field in Fig.~\ref{fig.kgmodel}(f), using the same model parameters  used to show the different phases in panel (e).
Note that $|\kappa_{xy}|/T$ first increases monotonically in the ZZ1 phase, stable below field $h_{c1}$, before changing sign in the ZZ2 phase.
In the region $h_{c2}<h<h_{c3}$, the computed magnon band structure becomes unphysical, meaning the failure of the zigzag ansatz to capture the true ground state, which may be a different four-spin order~\cite{Chern2021PRL}, or a magnetic order with an enlarged unit cell~\cite{Chern-YBK-semiclassics2020} that is beyond the scope of the present study -- we use UN to represent this unknown phase. Finally, the system enters fully polarized phase for fields $h>h_{c3}$, where $|\kappa_{xy}|/T$ decreases with increasing field. 
As this plot illustrates, the intensity of the thermal Hall effect thus obtained is always smaller than at most 0.2 in the fermionic units -- or about 40\% of the  value observed in Ref.~\cite{Matsuda-quantized}.
Below, we explore what the upper bound is on the thermal conductivity as a function of model parameters.


\vspace{1mm}
\textit{Upper bound on the thermal Hall effect due to magnons}.
Since the precise values of the model parameters in Eq.~\eqref{eq.model} corresponding to \rucl\  are still under intense debate (see Table 1 in Ref.~\cite{SashaC} for a list of different proposals), we scan a wide range of the physically relevant parameter values with the goal of determining the upper bound on $\kappa^{2D}_{xy}/T$. The parameter ranges we used are $0<\Gamma<7.2$~meV, $-4.2 <J_1<0$~meV, $0<J_3<4.0$ meV, and $-3.6<\Gamma'<3.6$ meV with step size $\delta=0.2$ meV (while keeping the magnitude of the Kitaev term $K=-7.2$~meV fixed  as stated earlier). 

\begin{figure}[!t]
    \centering
    \includegraphics[width=0.49\textwidth]{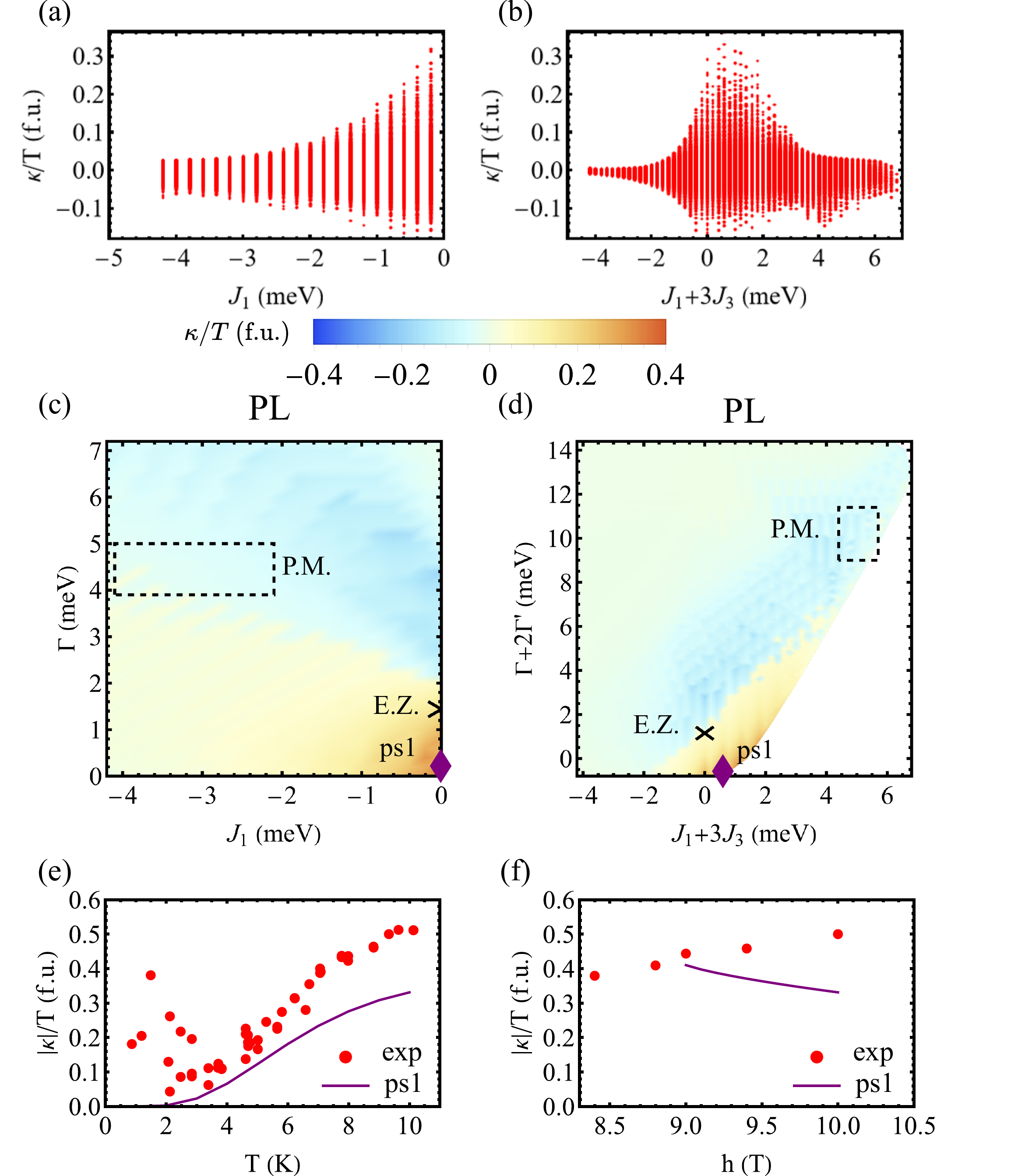}
    \caption{(a,b) The distribution of $\kappa_{xy}/T$ as a function of $J_1$ and $J_1+3J_3$ in polarized phase at $h=10$ T and $T=10$~K. (c,d) The largest value of  $\kappa_{xy}/T$ in PL phase  plotted as a false color map against (c) $J_1$, $\Gamma$ and (d) $J_1+3J_3$,  $\Gamma\!+\!2\Gamma'$. The region surrounded by the dashed line corresponds to the ``realistic parameter region'' in Ref.~\cite{SashaC}, the parameter set in Ref.~\cite{YBK} is labelled by the cross. (e) The temperature dependence (in $h\!=\!10$~T) and (f) magnetic field dependence (at $T\!=\!10$~K) of $\kappa_{xy}/T$ for the parameter set 1 (ps1): $\{K,J_1,J_3,\Gamma,\Gamma'\}=\{-7.2,0,0.2,0.2,-0.4\}$ meV, represented by the purple diamond in panels (c,d).
    }
    \vspace{-3mm}
    \label{fig.pl}
\end{figure}

According to the recent experimental data in Ref.~\cite{ONG}, the thermal Hall conductivity tends to be largest in the high field range $h\approx 10$~T and at moderately high temperatures $8\lesssim T\lesssim 12$~K. Therefore, for concreteness, we investigate the magnitude of $\kappa^{2D}_{xy}/T$ under the relevant experimental conditions $h=10$~T and $T=10$~K. 
We first start with the fully polarized (PL) phase. Because of the difficulty of representing the plots in the four-dimensional parameter space ($\Gamma,\Gamma',J_1,J_3$), we choose to plot the distribution of the $\kappa^{2D}_{xy}/T$ values vs. $J_1$ in   Fig.~{\ref{fig.pl}}(a), with each data point corresponding to a different choice o the remaining parameters. It is clear from this panel (a) that the largest values of $\kappa^{2D}_{xy}/T$ are attained at negative and very small $J_1$. This is because a weak ferromagnetic $J_1$ will destabilize the polarized state, and result in the lower magnon band moving down, which leads to stronger magnon modes contribution to the thermal Hall effect because of the increasing weight of the $c_2(x)$ function in Eq.~\eqref{eq.kxy}. In Ref.~\cite{SashaC} the authors identified the linear combination $J_1+3J_3$ as the relevant parameter, which is used as the $x$-axis to plot the calculated $\kappa^{2D}_{xy}/T$ in Fig.~\ref{fig.pl}(b), with similar conclusions reached.

An alternative way of looking at the data is to plot the maximum value of $\kappa^{2D}_{xy}/T$ as a false color on a two-dimensional plot with axes given by $J_1$ and $\Gamma$, which we do in Fig.~\ref{fig.pl}(c), or following the strategy proposed in Ref.~\cite{SashaC}, with the axes formed by effective couplings $\tilde{J} =J_1 + 3J_3$ and $\tilde{\Gamma}=\Gamma+2\Gamma'$, shown in Fig.~\ref{fig.pl}(d).
Each data point in these panels is taken to be the maximal value of $|\kappa^{2D}_{xy}|/T$
from varying the remaining parameters.
We find that the largest $|\kappa^{2D}_{xy}|/T$ in the PL phase (at $h=10$~T and $T=10$~K) never exceeds about 0.35 f.u. in the fermionic units. To orient the reader, we show with the dashed rectangle what the authors of  Ref.~\cite{SashaC} call the ``realistic parameter regime,''  and the cross represents the parameter set chosen in Ref. \cite{YBK}. In both regions, we find $|\kappa^{2D}_{xy}|/T$  to be less than 0.2 f.u., far from 0.5 f.u. reported in Ref.~\cite{Matsuda-quantized} and much below the maximum value measured in Ref.~\cite{ONG}.

Finally, we select the parameter set (labeled ps1) with the largest value of thermal Hall effect in our data and plot its value as a function of temperature and field, shown in Figs.~\ref{fig.pl}(e,f). While the monotonically increasing temperature dependence observed in Ref.~\cite{ONG} is qualitatively reproduced, the field dependence is opposite -- the experiment shows an increasing $|\kappa_{xy}(h)|$, while our data invariably decrease monotonically. 
Its physical reason  is clear -- in the fully polarized phase, the increase of the magnetic field leads to the (linear in field) growth of the magnon gap $\epsilon_\text{min}$. 
And since the Berry curvature integrand in Eq.~\eqref{eq.kxy} is weighted by the function $c_2(f(\epsilon))\propto \exp(-\epsilon/T)$, its value is exponentially suppressed at the experimentally relevant temperatures, leading to the decrease in  $|\kappa_{xy}(h)|$. 
We thus conclude that not only is the magnon contribution too low to account for the experimental value of thermal Hall effect, but its field dependence in the fully polarized phase cannot reproduce the experiment neither.

\begin{figure}[!t]
    \centering
    \includegraphics[width=0.5\textwidth]{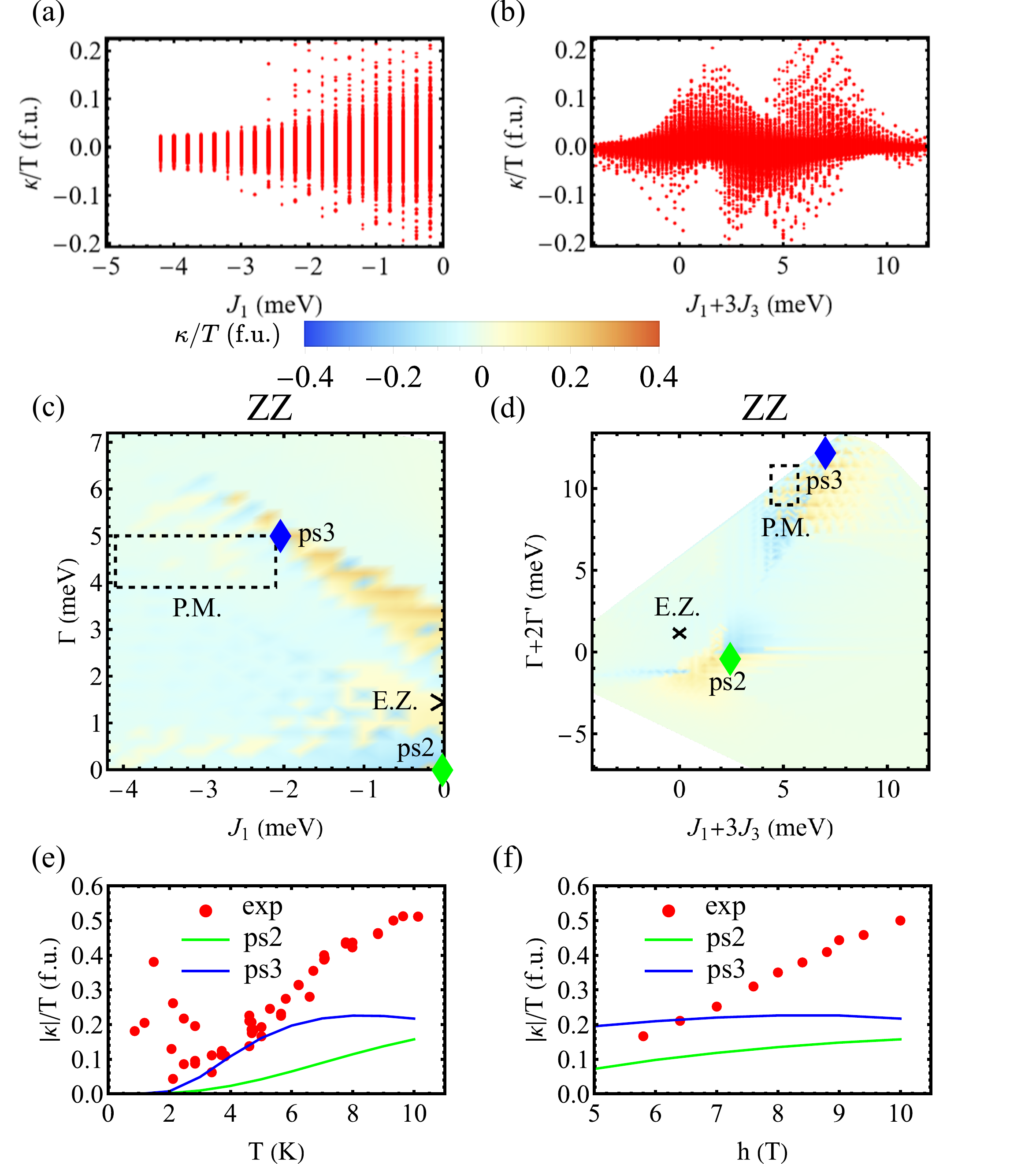}
\caption{(a,b)$\,$The distribution of $\kappa_{xy}/T$ as a function of $J_1$ and $J_1\!+\!3J_3$ in the zigzag phases at $h\!=\!10$ T and $T\!=\!10$~K. (c,d) The largest value of  $\kappa_{xy}/T$ in zigzag phases  plotted as a false color map against (c) $J_1$, $\Gamma$ and (d) $J_1\!+\!3J_3$,  $\Gamma\!+\!2\Gamma'$. The region surrounded by the dashed line corresponds to the ``realistic parameter region'' in Ref.~\cite{SashaC}, the parameter set in Ref.~\cite{YBK} is labelled by the cross. (e) The temperature and (f) magnetic field dependence of $\kappa_{xy}/T$ for the parameter sets
ps2 $\{K,J_1,J_3,\Gamma,\Gamma'\}=\{-7.2,0,0.8,0,-0.2\}$ meV and ps3 $\{K,J_1,J_3,\Gamma,\Gamma'\}=\{-7.2,-2,3,5,3.6\}$ meV, 
represented by the green and blue diamonds, respectively, in panels (c,d). 
    }
    \vspace{-3mm}
    \label{fig.zz}
\end{figure}

We therefore turn our attention to the zigzag phases, with the distributions of $\kappa_{xy}^{2D}/T$ along the $J_1$ and $\tilde{J}=J_1+3J_3$ shown in Figs.~\ref{fig.zz} (a) and (b), respectively, evaluated at $h=10$~T and $T=10$~K as before. 
Similarly to the PL phase, the magnitude of the Hall conductivity increases with the decreasing $|J_1|$. The maximum value of $\kappa^{2D}_{xy}/T$ is plotted as a false color map in $(J_1,\Gamma)$ and $(\tilde{J},\tilde{\Gamma})$ coordinates, respectively, in Figs.~\ref{fig.zz}(c) and    (d). 
We find the largest value of $\kappa^{2D}_{xy}/T$ to be about 0.2 f.u. When considering the ``realistic parameter regime'' in Ref. \cite{SashaC} (dashed region) or the parameters used in Ref.~\cite{YBK} (cross), $\kappa^{2D}_{xy}/T$ does not exceed 0.15 f.u., far below the experimentally reported values~\cite{Matsuda-quantized,Matsuda-quantized2,ONG}.

To compare with the experimental data, we choose two sets of parameters (labeled ps2 and ps3 in Fig.~\ref{fig.zz}, the values are listed in the caption) that belong to the ZZ1 and ZZ2 phases, respectively, and which have the largest values of $|\kappa_{xy}^{2D}|/T$ in our studied range. We evaluate the temperature and field dependence of $|\kappa^{2D}_{xy}|/T$ at these parameter sets and plot them against the experimental data from Ref.~\cite{ONG} in Figs.~\ref{fig.zz}(e,f).
Under the experimentally relevant conditions $2<T<6$~K and $6<h<7$~T, we find that the 
parameter set ps2 qualitatively matches the trends in the temperature and field dependence of the experimental thermal Hall data. 
Near $T=10$ K and $h=10$ T, $\kappa_{xy}/T$ continues to increase for ps2 (as is the case experimentally); whereas for ps3 it starts to decline. However, as noted already and as seen from Figs.~\ref{fig.zz}(e,f), the computed magnitude  of $|\kappa^{2D}_{xy}|/T$ is well below the experimental values. This indicates that the intrinsic magnon contribution alone cannot fully account for the measured thermal Hall conductivity in \rucl. We thus turn our attention to additional, bosonic in nature contributions to the thermal Hall effect.

\begin{figure}[!ht]
    \centering
    \includegraphics[width=0.48\textwidth]{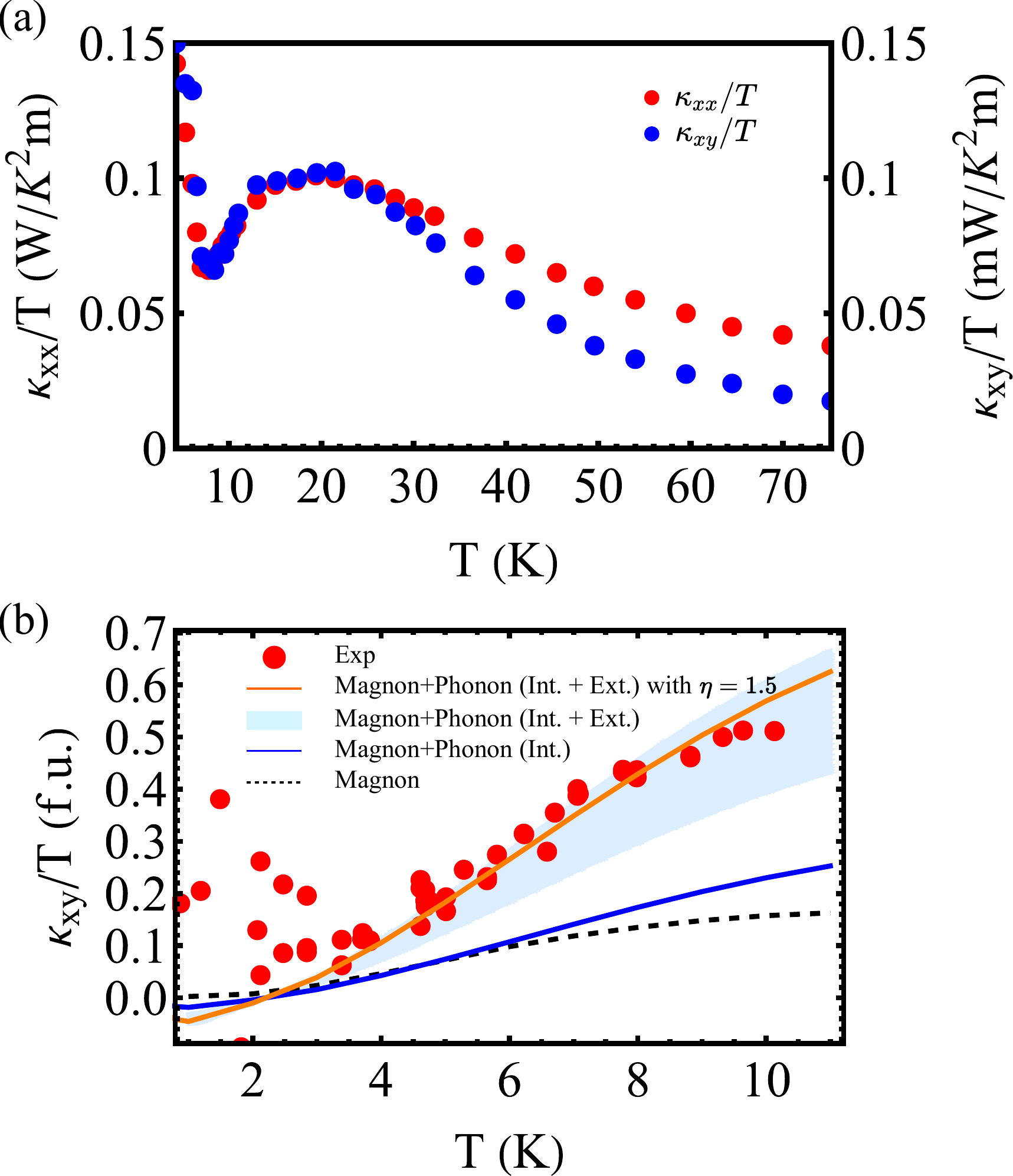}
    \caption{
    (a) The experimentally measured $\kappa_{xx}$ and $\kappa_{xy}$ in \rucl\ as functions of temperature, adopted from Ref.~\cite{taillefer-phonons2022}. 
    (b) Total computed $\kappa_{xy}/T$,  contributed from different sources, compared with the experimental data (red circles) from Ref.~\cite{ONG}.
    The magnon and phonon intrinsic component $\kappa^{I}_{xy}$ (solid line, for model ps2), summed together with the extrinsic phonon contribution $\kappa^{E}_{xy}$, is indicated with the blue shaded region (whose width is given by the experimental uncertainty in determining $\kappa^{E}_{xy}=\eta \kappa^{I}_{xy}$, see text). Selecting the value $\eta\!=\!1.5$ inside this region gives the best fit (orange line) to the experimental data for this particular model parameters (ps2).
    \label{fig.phonon}}
\end{figure}

\vspace{1mm}
\textit{Phonon contribution to the thermal Hall effect.}
We now turn to investigate other effects that may enhance the thermal Hall effect.
Recent experimental data~\cite{taillefer-phonons2022} on \rucl\, show that the temperature dependence of $\kappa_{xy}$ resembles closely that of longitudinal thermal conductivity $\kappa_{xx}$, as demonstrated in Fig.~\ref{fig.phonon}(a), with the ratio between the two roughly the same ($0.03-0.10$\%) across different samples, from which the authors of Ref.~\cite{taillefer-phonons2022} conclude that phonons must play a key role in the thermal Hall effect. 
In an insulator, the logitudinal $\kappa_{xx}$ is dominated by non-chiral acoustic phonons. The thermal Hall effect by contrast is time-reversal odd and chiral in nature. 
We distinguish two mechanisms 
of such chiral phonon response: 
the intrinsic one, due to the Berry curvature induced by magnon-phonon coupling, and the extrinsic one, due to phonon scattering off of defects.

\vspace{1mm}
\textit{Intrinsic phonon contribution to $\kappa_{xy}$.}
The distance dependence of the superexchange interactions between Ru$^{3+}$ ions leads naturally to the magnetoelastic (ME) coupling of the generic form (see SM for further details):
\vspace{-2mm}
\beq
\label{eq.me.interaction}
H_\text{ME} = \sum_{i,j} S_i^\alpha S_j^\beta\,  (\vec{u}_i - \vec{u}_j)\cdot\vec{{\nabla}}_{\mathbf{r}_{ij}}
\, J^{\alpha\beta}(\mathbf{r}_{ij}),
\eeq
where $\vec{u}_i$ is the displacement of the  ion at site $i$ from its equilibrium position. 
Writing these displacements in terms of the phonon operators
$u_i^\gamma\sim (a_{i\gamma}^\dagger + a_{i\gamma})$ (with polarization $\gamma$), this results in the hybridization between the magnons and the   phonon branches \cite{Kittel,LiePhysRevLett2021,LebertPhysRevB2022}, endowing the phonons with the Berry curvature that contributes to $\kappa_{xy}$ just as in Eq.~\eqref{eq.kxy}. 
The magnitude of the additional phonon contribution $\kappa_{xy}^\text{(I)}$ is related to the strength of the ME coupling,
whose existence is supported by the softening of the phonon branch at small $q$-vectors meansured in experiments~\cite{LiNatComm2021}. 
Choosing the  ME coupling that qualitatively reproduces the phonon softening (see SM for details), the value of $\kappa_{xy}/T$ is plotted in Fig.~\ref{fig.phonon}(b) with the solid blue line using the parameter set ps2 in the ZZ1 phase. As this figure shows, the magnitude of the intrinsic $\kappa^{I}_{xy}/T$ is enhanced compared to the magnon-only value in the temperature regime of interest $T>3$~K -- the same conclusion was reached in the recent study Ref.~\cite{Li-Okamoto2022}. However, the  magnitude $\kappa^{I}_{xy}/T$ ends up still being smaller than the experimentally measured Hall signal, prompting us to consider additional, phonon contributions.

\vspace{-2mm}
\textit{Extrinsic phonon contribution to $\kappa_{xy}$.}
%
There are multiple sources of phonon contributions to the thermal Hall effect,
in analogy to the phonon contribution to the electronic Hall effect ($\sigma_{xy}$) in metals~\cite{Nagaosa-RMP}.
One candidate mechanism is the \textit{intrinsic skew-scattering}, which originates from the Lorentz force on ions~\cite{Flebus-MacDonald2022}. 
Another source is the \textit{extrinsic skew-scattering} from phonons scattering off of magnetic impurities~\cite{Chen-Kivelson2020,Flebus-MacDonald2022}.
However, after comparing the experimental data~\cite{taillefer-phonons2022} with characteristic features of these effects (see SM for detailed analysis),
we came to the conclusion that they are negligible in \rucl.

Instead, the very weak temperature dependence of the ratio $\kappa_{xy}/\kappa_{xx}$ (see Fig.~\ref{fig.phonon}a) and absence of strong sample variability in recent experiments~\cite{taillefer-phonons2022} indicate that the dominant extrinsic contribution to $\kappa_{xy}$  is most likely from the so-called \textit{side-jump scattering} of phonons off of defects  \cite{Nagaosa-RMP},
which we demonstrate in the SM using the formalism recently developed in Ref.~\cite{Guo-Sachdev2022}.
Crucially, this effect scales with the phonon mean-free path $\ell$, just like the longitudinal thermal conductivity, consistent with the ratio of the two being sample independent. 

Using the experimental data from Ref.~\cite{taillefer-phonons2022} at high temperatures $T\gtrsim |K|/k_B \approx 80K$, above the magnon bandwidth where the effects of the Kitaev physics and associated Berry curvature are unimportant, we determine the Hall angle due to extrinsic scattering $\theta_H^E\equiv \kappa_{xy}^E/\kappa_{xx} = (0.6\pm0.2)\times 10^{-3}$. 
By contrast, at low temperatures $T\sim 10-15$~K where the interpretation of the thermal Hall measurements on \rucl\  is disputed, both the intrinsic (due to the Berry curvature) and extrinsic contributions to $\kappa_{xy}$ must be taken into account. By comparing the data in this low-$T$ region with the value $\theta_H^E$ from above, we are able to determine the phenomenological ratio $\eta\equiv \kappa_{xy}^E/\kappa_{xy}^I$. We obtain (see SM for details) $\eta=1.2\pm 0.5$ with the uncertainty related to the spread of the experimental data among the different samples. 
Taking the intrinsic $\kappa_{xy}^I$ (solid line in Fig.~\ref{fig.phonon}(b)) and multiplying by $(1+\eta)$, we thus obtain the total $\kappa_{xy}=\kappa_{xy}^I + \kappa_{xy}^E$, marked by the shaded blue region in Fig.~\ref{fig.phonon}(b). 
The experimental data from Ong's group~\cite{ONG} fall 
inside the yellow shaded region.
This indicates that the   thermal Hall effect 
from the bosonic model given our parameter choice can explain the experimental observation.

Furthermore, given the flexibility in  the magnitude of the ME coupling, we show (see SM) that choosing the largest physically allowed coupling results in the intrinsic value of $\kappa_{xy}^{I}/T$ of the order of 0.35 f.u. (for ps2), and further including the extrinsic effects can yield values of thermal Hall effect even in excess of the experimentally measured values. We thus conclude that the magnon-phonon mechanism proposed in this work is more than sufficient to explain the experimental thermal Hall data in \rucl~ without overly fine-tuning the model parameters, i.e., in a finite region in the parameter space.

\vspace{1mm}
\textit{Discussion.}
Having scanned a broad range of physically motivated parameters of the $(K,J_1,J_3,\Gamma,\Gamma')$ generalized Kitaev--Heisenberg model, we conclude that the intrinsic magnon contribution alone is insufficient to explain the large observed magnitude of the thermal Hall effect in \rucl. We further found that in order to reconcile the observed magnetic field dependence of the thermal Hall effect, it is necessary to conclude that \rucl remains in the canted zigzag phase (as opposed to field polarized) in fields up to $\sim 10$~T. 
This conclusion is supported by the recent study by Li and Okamoto~\cite{Li-Okamoto2022} who found that the spin-phonon coupling tends to stabilize the canted zigzag phase for higher applied fields, compared to the pure spin model which would otherwise become fully polarized.
Taking into account the spin-phonon coupling endows phonons with the chirality, contributing an additional intrinsic term to $\kappa_{xy}^{I}$, which however still falls short of explaining the experimental data, necessitating the inclusion of an extrinsic source of Hall effect. 
With minimal assumptions as to its mechanism, we used the existing experimental data to quantitatively arrive at the measure  $\kappa_{xy}^E/\kappa_{xy}^I$ yielding  values between 1 and 2, meaning that the extrinsic phonon contribution to $\kappa_{xy}$ is comparable, or a little larger, than the intrinsic Berry curvature effect. Taking both into account, we are able to explain not only the large magnitude but also the detailed temperature dependence of $\kappa_{xy}$, which is bosonic in nature.

\vspace{1mm}
\textit{Acknowledgements.} The authors thank P. Ong, I. Sodemann and S. Winter for fruitful discussions, and L. Taillefer for the critical reading of the manuscript. H.Y. and A.H.N. were supported by the National Science Foundation Division of Materials Research under the Award DMR-1917511. S.L. was supported by the Robert A. Welch Foundation grant No.~C-1818.

\bibliography{cite}

\begin{thebibliography}{49}%
\makeatletter
\providecommand \@ifxundefined [1]{%
 \@ifx{#1\undefined}
}%
\providecommand \@ifnum [1]{%
 \ifnum #1\expandafter \@firstoftwo
 \else \expandafter \@secondoftwo
 \fi
}%
\providecommand \@ifx [1]{%
 \ifx #1\expandafter \@firstoftwo
 \else \expandafter \@secondoftwo
 \fi
}%
\providecommand \natexlab [1]{#1}%
\providecommand \enquote  [1]{``#1''}%
\providecommand \bibnamefont  [1]{#1}%
\providecommand \bibfnamefont [1]{#1}%
\providecommand \citenamefont [1]{#1}%
\providecommand \href@noop [0]{\@secondoftwo}%
\providecommand \href [0]{\begingroup \@sanitize@url \@href}%
\providecommand \@href[1]{\@@startlink{#1}\@@href}%
\providecommand \@@href[1]{\endgroup#1\@@endlink}%
\providecommand \@sanitize@url [0]{\catcode `\\12\catcode `\$12\catcode
  `\&12\catcode `\#12\catcode `\^12\catcode `\_12\catcode `\%12\relax}%
\providecommand \@@startlink[1]{}%
\providecommand \@@endlink[0]{}%
\providecommand \url  [0]{\begingroup\@sanitize@url \@url }%
\providecommand \@url [1]{\endgroup\@href {#1}{\urlprefix }}%
\providecommand \urlprefix  [0]{URL }%
\providecommand \Eprint [0]{\href }%
\providecommand \doibase [0]{http://dx.doi.org/}%
\providecommand \selectlanguage [0]{\@gobble}%
\providecommand \bibinfo  [0]{\@secondoftwo}%
\providecommand \bibfield  [0]{\@secondoftwo}%
\providecommand \translation [1]{[#1]}%
\providecommand \BibitemOpen [0]{}%
\providecommand \bibitemStop [0]{}%
\providecommand \bibitemNoStop [0]{.\EOS\space}%
\providecommand \EOS [0]{\spacefactor3000\relax}%
\providecommand \BibitemShut  [1]{\csname bibitem#1\endcsname}%
\let\auto@bib@innerbib\@empty
\bibitem [{\citenamefont {Plumb}\ \emph {et~al.}(2014)\citenamefont {Plumb},
  \citenamefont {Clancy}, \citenamefont {Sandilands}, \citenamefont {Shankar},
  \citenamefont {Hu}, \citenamefont {Burch}, \citenamefont {Kee},\ and\
  \citenamefont {Kim}}]{Plumb2014}%
  \BibitemOpen
  \bibfield  {author} {\bibinfo {author} {\bibfnamefont {K.~W.}\ \bibnamefont
  {Plumb}}, \bibinfo {author} {\bibfnamefont {J.~P.}\ \bibnamefont {Clancy}},
  \bibinfo {author} {\bibfnamefont {L.~J.}\ \bibnamefont {Sandilands}},
  \bibinfo {author} {\bibfnamefont {V.~V.}\ \bibnamefont {Shankar}}, \bibinfo
  {author} {\bibfnamefont {Y.~F.}\ \bibnamefont {Hu}}, \bibinfo {author}
  {\bibfnamefont {K.~S.}\ \bibnamefont {Burch}}, \bibinfo {author}
  {\bibfnamefont {H.-Y.}\ \bibnamefont {Kee}}, \ and\ \bibinfo {author}
  {\bibfnamefont {Y.-J.}\ \bibnamefont {Kim}},\ }\href {\doibase
  10.1103/PhysRevB.90.041112} {\bibfield  {journal} {\bibinfo  {journal} {Phys.
  Rev. B}\ }\textbf {\bibinfo {volume} {90}},\ \bibinfo {pages} {041112}
  (\bibinfo {year} {2014})}\BibitemShut {NoStop}%
\bibitem [{\citenamefont {Kitaev}(2006)}]{kitaev2006}%
  \BibitemOpen
  \bibfield  {author} {\bibinfo {author} {\bibfnamefont {A.}~\bibnamefont
  {Kitaev}},\ }\href@noop {} {\bibfield  {journal} {\bibinfo  {journal} {Annals
  of Physics}\ }\textbf {\bibinfo {volume} {321}},\ \bibinfo {pages} {2}
  (\bibinfo {year} {2006})}\BibitemShut {NoStop}%
\bibitem [{\citenamefont {Cao}\ \emph {et~al.}(2016)\citenamefont {Cao},
  \citenamefont {Banerjee}, \citenamefont {Yan}, \citenamefont {Bridges},
  \citenamefont {Lumsden}, \citenamefont {Mandrus}, \citenamefont {Tennant},
  \citenamefont {Chakoumakos},\ and\ \citenamefont {Nagler}}]{cao2016}%
  \BibitemOpen
  \bibfield  {author} {\bibinfo {author} {\bibfnamefont {H.~B.}\ \bibnamefont
  {Cao}}, \bibinfo {author} {\bibfnamefont {A.}~\bibnamefont {Banerjee}},
  \bibinfo {author} {\bibfnamefont {J.-Q.}\ \bibnamefont {Yan}}, \bibinfo
  {author} {\bibfnamefont {C.~A.}\ \bibnamefont {Bridges}}, \bibinfo {author}
  {\bibfnamefont {M.~D.}\ \bibnamefont {Lumsden}}, \bibinfo {author}
  {\bibfnamefont {D.~G.}\ \bibnamefont {Mandrus}}, \bibinfo {author}
  {\bibfnamefont {D.~A.}\ \bibnamefont {Tennant}}, \bibinfo {author}
  {\bibfnamefont {B.~C.}\ \bibnamefont {Chakoumakos}}, \ and\ \bibinfo {author}
  {\bibfnamefont {S.~E.}\ \bibnamefont {Nagler}},\ }\href {\doibase
  10.1103/PhysRevB.93.134423} {\bibfield  {journal} {\bibinfo  {journal} {Phys.
  Rev. B}\ }\textbf {\bibinfo {volume} {93}},\ \bibinfo {pages} {134423}
  (\bibinfo {year} {2016})}\BibitemShut {NoStop}%
\bibitem [{\citenamefont {Kasahara}\ \emph {et~al.}(2018)\citenamefont
  {Kasahara}, \citenamefont {Ohnishi}, \citenamefont {Mizukami}, \citenamefont
  {Tanaka}, \citenamefont {Ma}, \citenamefont {Sugii}, \citenamefont {Kurita},
  \citenamefont {Tanaka}, \citenamefont {Nasu}, \citenamefont {Motome},
  \citenamefont {Shibauchi},\ and\ \citenamefont
  {Matsuda}}]{Matsuda-quantized}%
  \BibitemOpen
  \bibfield  {author} {\bibinfo {author} {\bibfnamefont {Y.}~\bibnamefont
  {Kasahara}}, \bibinfo {author} {\bibfnamefont {T.}~\bibnamefont {Ohnishi}},
  \bibinfo {author} {\bibfnamefont {Y.}~\bibnamefont {Mizukami}}, \bibinfo
  {author} {\bibfnamefont {O.}~\bibnamefont {Tanaka}}, \bibinfo {author}
  {\bibfnamefont {S.}~\bibnamefont {Ma}}, \bibinfo {author} {\bibfnamefont
  {K.}~\bibnamefont {Sugii}}, \bibinfo {author} {\bibfnamefont
  {N.}~\bibnamefont {Kurita}}, \bibinfo {author} {\bibfnamefont
  {H.}~\bibnamefont {Tanaka}}, \bibinfo {author} {\bibfnamefont
  {J.}~\bibnamefont {Nasu}}, \bibinfo {author} {\bibfnamefont {Y.}~\bibnamefont
  {Motome}}, \bibinfo {author} {\bibfnamefont {T.}~\bibnamefont {Shibauchi}}, \
  and\ \bibinfo {author} {\bibfnamefont {Y.}~\bibnamefont {Matsuda}},\ }\href
  {\doibase 10.1038/s41586-018-0274-0} {\bibfield  {journal} {\bibinfo
  {journal} {Nature}\ }\textbf {\bibinfo {volume} {559}},\ \bibinfo {pages}
  {227} (\bibinfo {year} {2018})}\BibitemShut {NoStop}%
\bibitem [{\citenamefont {Yokoi}\ \emph {et~al.}(2021)\citenamefont {Yokoi},
  \citenamefont {Ma}, \citenamefont {Kasahara}, \citenamefont {Kasahara},
  \citenamefont {Shibauchi}, \citenamefont {Kurita}, \citenamefont {Tanaka},
  \citenamefont {Nasu}, \citenamefont {Motome}, \citenamefont {Hickey},
  \citenamefont {Trebst},\ and\ \citenamefont {Matsuda}}]{Matsuda-quantized2}%
  \BibitemOpen
  \bibfield  {author} {\bibinfo {author} {\bibfnamefont {T.}~\bibnamefont
  {Yokoi}}, \bibinfo {author} {\bibfnamefont {S.}~\bibnamefont {Ma}}, \bibinfo
  {author} {\bibfnamefont {Y.}~\bibnamefont {Kasahara}}, \bibinfo {author}
  {\bibfnamefont {S.}~\bibnamefont {Kasahara}}, \bibinfo {author}
  {\bibfnamefont {T.}~\bibnamefont {Shibauchi}}, \bibinfo {author}
  {\bibfnamefont {N.}~\bibnamefont {Kurita}}, \bibinfo {author} {\bibfnamefont
  {H.}~\bibnamefont {Tanaka}}, \bibinfo {author} {\bibfnamefont
  {J.}~\bibnamefont {Nasu}}, \bibinfo {author} {\bibfnamefont {Y.}~\bibnamefont
  {Motome}}, \bibinfo {author} {\bibfnamefont {C.}~\bibnamefont {Hickey}},
  \bibinfo {author} {\bibfnamefont {S.}~\bibnamefont {Trebst}}, \ and\ \bibinfo
  {author} {\bibfnamefont {Y.}~\bibnamefont {Matsuda}},\ }\href {\doibase
  10.1126/science.aay5551} {\bibfield  {journal} {\bibinfo  {journal}
  {Science}\ }\textbf {\bibinfo {volume} {373}},\ \bibinfo {pages} {568}
  (\bibinfo {year} {2021})}\BibitemShut {NoStop}%
\bibitem [{\citenamefont {Nasu}\ \emph {et~al.}(2017)\citenamefont {Nasu},
  \citenamefont {Yoshitake},\ and\ \citenamefont {Motome}}]{Nasu2017}%
  \BibitemOpen
  \bibfield  {author} {\bibinfo {author} {\bibfnamefont {J.}~\bibnamefont
  {Nasu}}, \bibinfo {author} {\bibfnamefont {J.}~\bibnamefont {Yoshitake}}, \
  and\ \bibinfo {author} {\bibfnamefont {Y.}~\bibnamefont {Motome}},\ }\href
  {\doibase 10.1103/PhysRevLett.119.127204} {\bibfield  {journal} {\bibinfo
  {journal} {Phys. Rev. Lett.}\ }\textbf {\bibinfo {volume} {119}},\ \bibinfo
  {pages} {127204} (\bibinfo {year} {2017})}\BibitemShut {NoStop}%
\bibitem [{\citenamefont {Czajka}\ \emph {et~al.}(2022)\citenamefont {Czajka},
  \citenamefont {Gao}, \citenamefont {Hirschberger}, \citenamefont
  {Lampen-Kelley}, \citenamefont {Banerjee}, \citenamefont {Quirk},
  \citenamefont {Mandrus}, \citenamefont {Nagler},\ and\ \citenamefont
  {Ong}}]{ONG}%
  \BibitemOpen
  \bibfield  {author} {\bibinfo {author} {\bibfnamefont {P.}~\bibnamefont
  {Czajka}}, \bibinfo {author} {\bibfnamefont {T.}~\bibnamefont {Gao}},
  \bibinfo {author} {\bibfnamefont {M.}~\bibnamefont {Hirschberger}}, \bibinfo
  {author} {\bibfnamefont {P.}~\bibnamefont {Lampen-Kelley}}, \bibinfo {author}
  {\bibfnamefont {A.}~\bibnamefont {Banerjee}}, \bibinfo {author}
  {\bibfnamefont {N.}~\bibnamefont {Quirk}}, \bibinfo {author} {\bibfnamefont
  {D.~G.}\ \bibnamefont {Mandrus}}, \bibinfo {author} {\bibfnamefont {S.~E.}\
  \bibnamefont {Nagler}}, \ and\ \bibinfo {author} {\bibfnamefont {N.~P.}\
  \bibnamefont {Ong}},\ }\href {\doibase 10.1038/s41563-022-01397-w} {\bibfield
   {journal} {\bibinfo  {journal} {Nature Materials}\ }\textbf {\bibinfo
  {volume} {22}},\ \bibinfo {pages} {36} (\bibinfo {year} {2022})}\BibitemShut
  {NoStop}%
\bibitem [{\citenamefont {McClarty}\ \emph {et~al.}(2018)\citenamefont
  {McClarty}, \citenamefont {Dong}, \citenamefont {Gohlke}, \citenamefont
  {Rau}, \citenamefont {Pollmann}, \citenamefont {Moessner},\ and\
  \citenamefont {Penc}}]{Mcclarty2018}%
  \BibitemOpen
  \bibfield  {author} {\bibinfo {author} {\bibfnamefont {P.~A.}\ \bibnamefont
  {McClarty}}, \bibinfo {author} {\bibfnamefont {X.-Y.}\ \bibnamefont {Dong}},
  \bibinfo {author} {\bibfnamefont {M.}~\bibnamefont {Gohlke}}, \bibinfo
  {author} {\bibfnamefont {J.~G.}\ \bibnamefont {Rau}}, \bibinfo {author}
  {\bibfnamefont {F.}~\bibnamefont {Pollmann}}, \bibinfo {author}
  {\bibfnamefont {R.}~\bibnamefont {Moessner}}, \ and\ \bibinfo {author}
  {\bibfnamefont {K.}~\bibnamefont {Penc}},\ }\href {\doibase
  10.1103/PhysRevB.98.060404} {\bibfield  {journal} {\bibinfo  {journal} {Phys.
  Rev. B}\ }\textbf {\bibinfo {volume} {98}},\ \bibinfo {pages} {060404}
  (\bibinfo {year} {2018})}\BibitemShut {NoStop}%
\bibitem [{\citenamefont {Cookmeyer}\ and\ \citenamefont
  {Moore}(2018)}]{CookmeyerPRB2018}%
  \BibitemOpen
  \bibfield  {author} {\bibinfo {author} {\bibfnamefont {T.}~\bibnamefont
  {Cookmeyer}}\ and\ \bibinfo {author} {\bibfnamefont {J.~E.}\ \bibnamefont
  {Moore}},\ }\href {\doibase 10.1103/PhysRevB.98.060412} {\bibfield  {journal}
  {\bibinfo  {journal} {Phys. Rev. B}\ }\textbf {\bibinfo {volume} {98}},\
  \bibinfo {pages} {060412} (\bibinfo {year} {2018})}\BibitemShut {NoStop}%
\bibitem [{\citenamefont {Chern}\ \emph {et~al.}(2021)\citenamefont {Chern},
  \citenamefont {Zhang},\ and\ \citenamefont {Kim}}]{Chern2021PRL}%
  \BibitemOpen
  \bibfield  {author} {\bibinfo {author} {\bibfnamefont {L.~E.}\ \bibnamefont
  {Chern}}, \bibinfo {author} {\bibfnamefont {E.~Z.}\ \bibnamefont {Zhang}}, \
  and\ \bibinfo {author} {\bibfnamefont {Y.~B.}\ \bibnamefont {Kim}},\ }\href
  {\doibase 10.1103/PhysRevLett.126.147201} {\bibfield  {journal} {\bibinfo
  {journal} {Phys. Rev. Lett.}\ }\textbf {\bibinfo {volume} {126}},\ \bibinfo
  {pages} {147201} (\bibinfo {year} {2021})}\BibitemShut {NoStop}%
\bibitem [{\citenamefont {Zhang}\ \emph {et~al.}(2021)\citenamefont {Zhang},
  \citenamefont {Chern},\ and\ \citenamefont {Kim}}]{YBK}%
  \BibitemOpen
  \bibfield  {author} {\bibinfo {author} {\bibfnamefont {E.~Z.}\ \bibnamefont
  {Zhang}}, \bibinfo {author} {\bibfnamefont {L.~E.}\ \bibnamefont {Chern}}, \
  and\ \bibinfo {author} {\bibfnamefont {Y.~B.}\ \bibnamefont {Kim}},\ }\href
  {\doibase 10.1103/PhysRevB.103.174402} {\bibfield  {journal} {\bibinfo
  {journal} {Phys. Rev. B}\ }\textbf {\bibinfo {volume} {103}},\ \bibinfo
  {pages} {174402} (\bibinfo {year} {2021})}\BibitemShut {NoStop}%
\bibitem [{\citenamefont {Li}\ and\ \citenamefont
  {Okamoto}(2022)}]{Li-Okamoto2022}%
  \BibitemOpen
  \bibfield  {author} {\bibinfo {author} {\bibfnamefont {S.}~\bibnamefont
  {Li}}\ and\ \bibinfo {author} {\bibfnamefont {S.}~\bibnamefont {Okamoto}},\
  }\href {\doibase 10.1103/PhysRevB.106.024413} {\bibfield  {journal} {\bibinfo
   {journal} {Phys. Rev. B}\ }\textbf {\bibinfo {volume} {106}},\ \bibinfo
  {pages} {024413} (\bibinfo {year} {2022})}\BibitemShut {NoStop}%
\bibitem [{\citenamefont {Bruin}\ \emph {et~al.}(2022)\citenamefont {Bruin},
  \citenamefont {Claus}, \citenamefont {Matsumoto}, \citenamefont {Kurita},
  \citenamefont {Tanaka},\ and\ \citenamefont {Takagi}}]{Takagi-quantized2022}%
  \BibitemOpen
  \bibfield  {author} {\bibinfo {author} {\bibfnamefont {J.~A.~N.}\
  \bibnamefont {Bruin}}, \bibinfo {author} {\bibfnamefont {R.~R.}\ \bibnamefont
  {Claus}}, \bibinfo {author} {\bibfnamefont {Y.}~\bibnamefont {Matsumoto}},
  \bibinfo {author} {\bibfnamefont {N.}~\bibnamefont {Kurita}}, \bibinfo
  {author} {\bibfnamefont {H.}~\bibnamefont {Tanaka}}, \ and\ \bibinfo {author}
  {\bibfnamefont {H.}~\bibnamefont {Takagi}},\ }\href {\doibase
  10.1038/s41567-021-01501-y} {\bibfield  {journal} {\bibinfo  {journal}
  {Nature Physics}\ }\textbf {\bibinfo {volume} {18}},\ \bibinfo {pages} {401}
  (\bibinfo {year} {2022})}\BibitemShut {NoStop}%
\bibitem [{\citenamefont {Lefran{\c{c}}ois}\ \emph {et~al.}(2022)\citenamefont
  {Lefran{\c{c}}ois}, \citenamefont {Grissonnanche}, \citenamefont {Baglo},
  \citenamefont {Lampen-Kelley}, \citenamefont {Yan}, \citenamefont {Balz},
  \citenamefont {Mandrus}, \citenamefont {Nagler}, \citenamefont {Kim},
  \citenamefont {Kim} \emph {et~al.}}]{taillefer-phonons2022}%
  \BibitemOpen
  \bibfield  {author} {\bibinfo {author} {\bibfnamefont {{\'E}.}~\bibnamefont
  {Lefran{\c{c}}ois}}, \bibinfo {author} {\bibfnamefont {G.}~\bibnamefont
  {Grissonnanche}}, \bibinfo {author} {\bibfnamefont {J.}~\bibnamefont
  {Baglo}}, \bibinfo {author} {\bibfnamefont {P.}~\bibnamefont
  {Lampen-Kelley}}, \bibinfo {author} {\bibfnamefont {J.-Q.}\ \bibnamefont
  {Yan}}, \bibinfo {author} {\bibfnamefont {C.}~\bibnamefont {Balz}}, \bibinfo
  {author} {\bibfnamefont {D.}~\bibnamefont {Mandrus}}, \bibinfo {author}
  {\bibfnamefont {S.}~\bibnamefont {Nagler}}, \bibinfo {author} {\bibfnamefont
  {S.}~\bibnamefont {Kim}}, \bibinfo {author} {\bibfnamefont {Y.-J.}\
  \bibnamefont {Kim}},  \emph {et~al.},\ }\href@noop {} {\bibfield  {journal}
  {\bibinfo  {journal} {Physical Review X}\ }\textbf {\bibinfo {volume} {12}},\
  \bibinfo {pages} {021025} (\bibinfo {year} {2022})}\BibitemShut {NoStop}%
\bibitem [{\citenamefont {Jackeli}\ and\ \citenamefont
  {Khaliullin}(2009)}]{Jackeli2009}%
  \BibitemOpen
  \bibfield  {author} {\bibinfo {author} {\bibfnamefont {G.}~\bibnamefont
  {Jackeli}}\ and\ \bibinfo {author} {\bibfnamefont {G.}~\bibnamefont
  {Khaliullin}},\ }\href {\doibase 10.1103/PhysRevLett.102.017205} {\bibfield
  {journal} {\bibinfo  {journal} {Phys. Rev. Lett.}\ }\textbf {\bibinfo
  {volume} {102}},\ \bibinfo {pages} {017205} (\bibinfo {year}
  {2009})}\BibitemShut {NoStop}%
\bibitem [{\citenamefont {Rau}\ \emph {et~al.}(2014)\citenamefont {Rau},
  \citenamefont {Lee},\ and\ \citenamefont {Kee}}]{Rau2014}%
  \BibitemOpen
  \bibfield  {author} {\bibinfo {author} {\bibfnamefont {J.~G.}\ \bibnamefont
  {Rau}}, \bibinfo {author} {\bibfnamefont {E.~K.-H.}\ \bibnamefont {Lee}}, \
  and\ \bibinfo {author} {\bibfnamefont {H.-Y.}\ \bibnamefont {Kee}},\ }\href
  {\doibase 10.1103/PhysRevLett.112.077204} {\bibfield  {journal} {\bibinfo
  {journal} {Phys. Rev. Lett.}\ }\textbf {\bibinfo {volume} {112}},\ \bibinfo
  {pages} {077204} (\bibinfo {year} {2014})}\BibitemShut {NoStop}%
\bibitem [{\citenamefont {Kim}\ \emph {et~al.}(2015)\citenamefont {Kim},
  \citenamefont {V.}, \citenamefont {Catuneanu},\ and\ \citenamefont
  {Kee}}]{Kim2015}%
  \BibitemOpen
  \bibfield  {author} {\bibinfo {author} {\bibfnamefont {H.-S.}\ \bibnamefont
  {Kim}}, \bibinfo {author} {\bibfnamefont {V.~S.}\ \bibnamefont {V.}},
  \bibinfo {author} {\bibfnamefont {A.}~\bibnamefont {Catuneanu}}, \ and\
  \bibinfo {author} {\bibfnamefont {H.-Y.}\ \bibnamefont {Kee}},\ }\href
  {\doibase 10.1103/PhysRevB.91.241110} {\bibfield  {journal} {\bibinfo
  {journal} {Phys. Rev. B}\ }\textbf {\bibinfo {volume} {91}},\ \bibinfo
  {pages} {241110} (\bibinfo {year} {2015})}\BibitemShut {NoStop}%
\bibitem [{\citenamefont {Kim}\ and\ \citenamefont {Kee}(2016)}]{Kim2016}%
  \BibitemOpen
  \bibfield  {author} {\bibinfo {author} {\bibfnamefont {H.-S.}\ \bibnamefont
  {Kim}}\ and\ \bibinfo {author} {\bibfnamefont {H.-Y.}\ \bibnamefont {Kee}},\
  }\href {\doibase 10.1103/PhysRevB.93.155143} {\bibfield  {journal} {\bibinfo
  {journal} {Phys. Rev. B}\ }\textbf {\bibinfo {volume} {93}},\ \bibinfo
  {pages} {155143} (\bibinfo {year} {2016})}\BibitemShut {NoStop}%
\bibitem [{\citenamefont {Winter}\ \emph {et~al.}(2016)\citenamefont {Winter},
  \citenamefont {Li}, \citenamefont {Jeschke},\ and\ \citenamefont
  {Valentí}}]{Winter2016}%
  \BibitemOpen
  \bibfield  {author} {\bibinfo {author} {\bibfnamefont {S.~M.}\ \bibnamefont
  {Winter}}, \bibinfo {author} {\bibfnamefont {Y.}~\bibnamefont {Li}}, \bibinfo
  {author} {\bibfnamefont {H.~O.}\ \bibnamefont {Jeschke}}, \ and\ \bibinfo
  {author} {\bibfnamefont {R.}~\bibnamefont {Valentí}},\ }\href {\doibase
  10.1103/PhysRevB.93.214431} {\bibfield  {journal} {\bibinfo  {journal} {Phys.
  Rev. B}\ }\textbf {\bibinfo {volume} {93}},\ \bibinfo {pages} {214431}
  (\bibinfo {year} {2016})}\BibitemShut {NoStop}%
\bibitem [{\citenamefont {Chaloupka}\ and\ \citenamefont
  {Khaliullin}(2016)}]{Chaloupka2016}%
  \BibitemOpen
  \bibfield  {author} {\bibinfo {author} {\bibfnamefont {J.}~\bibnamefont
  {Chaloupka}}\ and\ \bibinfo {author} {\bibfnamefont {G.}~\bibnamefont
  {Khaliullin}},\ }\href {\doibase 10.1103/PhysRevB.94.064435} {\bibfield
  {journal} {\bibinfo  {journal} {Phys. Rev. B}\ }\textbf {\bibinfo {volume}
  {94}},\ \bibinfo {pages} {064435} (\bibinfo {year} {2016})}\BibitemShut
  {NoStop}%
\bibitem [{\citenamefont {Yadav}\ \emph {et~al.}(2016)\citenamefont {Yadav},
  \citenamefont {Bogdanov}, \citenamefont {Katukuri}, \citenamefont
  {Nishimoto}, \citenamefont {Van Den~Brink},\ and\ \citenamefont
  {Hozoi}}]{Yadav2016}%
  \BibitemOpen
  \bibfield  {author} {\bibinfo {author} {\bibfnamefont {R.}~\bibnamefont
  {Yadav}}, \bibinfo {author} {\bibfnamefont {N.~A.}\ \bibnamefont {Bogdanov}},
  \bibinfo {author} {\bibfnamefont {V.~M.}\ \bibnamefont {Katukuri}}, \bibinfo
  {author} {\bibfnamefont {S.}~\bibnamefont {Nishimoto}}, \bibinfo {author}
  {\bibfnamefont {J.}~\bibnamefont {Van Den~Brink}}, \ and\ \bibinfo {author}
  {\bibfnamefont {L.}~\bibnamefont {Hozoi}},\ }\href@noop {} {\bibfield
  {journal} {\bibinfo  {journal} {Scientific reports}\ }\textbf {\bibinfo
  {volume} {6}},\ \bibinfo {pages} {1} (\bibinfo {year} {2016})}\BibitemShut
  {NoStop}%
\bibitem [{\citenamefont {Winter}\ \emph {et~al.}(2017)\citenamefont {Winter},
  \citenamefont {Tsirlin}, \citenamefont {Daghofer}, \citenamefont {van~den
  Brink}, \citenamefont {Singh}, \citenamefont {Gegenwart},\ and\ \citenamefont
  {Valent{\'{\i}}}}]{Winter2017}%
  \BibitemOpen
  \bibfield  {author} {\bibinfo {author} {\bibfnamefont {S.~M.}\ \bibnamefont
  {Winter}}, \bibinfo {author} {\bibfnamefont {A.~A.}\ \bibnamefont {Tsirlin}},
  \bibinfo {author} {\bibfnamefont {M.}~\bibnamefont {Daghofer}}, \bibinfo
  {author} {\bibfnamefont {J.}~\bibnamefont {van~den Brink}}, \bibinfo {author}
  {\bibfnamefont {Y.}~\bibnamefont {Singh}}, \bibinfo {author} {\bibfnamefont
  {P.}~\bibnamefont {Gegenwart}}, \ and\ \bibinfo {author} {\bibfnamefont
  {R.}~\bibnamefont {Valent{\'{\i}}}},\ }\href {\doibase
  10.1088/1361-648x/aa8cf5} {\bibfield  {journal} {\bibinfo  {journal} {Journal
  of Physics: Condensed Matter}\ }\textbf {\bibinfo {volume} {29}},\ \bibinfo
  {pages} {493002} (\bibinfo {year} {2017})}\BibitemShut {NoStop}%
\bibitem [{\citenamefont {Hou}\ \emph {et~al.}(2017)\citenamefont {Hou},
  \citenamefont {Xiang},\ and\ \citenamefont {Gong}}]{Hou2017}%
  \BibitemOpen
  \bibfield  {author} {\bibinfo {author} {\bibfnamefont {Y.~S.}\ \bibnamefont
  {Hou}}, \bibinfo {author} {\bibfnamefont {H.~J.}\ \bibnamefont {Xiang}}, \
  and\ \bibinfo {author} {\bibfnamefont {X.~G.}\ \bibnamefont {Gong}},\ }\href
  {\doibase 10.1103/PhysRevB.96.054410} {\bibfield  {journal} {\bibinfo
  {journal} {Phys. Rev. B}\ }\textbf {\bibinfo {volume} {96}},\ \bibinfo
  {pages} {054410} (\bibinfo {year} {2017})}\BibitemShut {NoStop}%
\bibitem [{\citenamefont {Winter}\ \emph {et~al.}(2018)\citenamefont {Winter},
  \citenamefont {Riedl}, \citenamefont {Kaib}, \citenamefont {Coldea},\ and\
  \citenamefont {Valentí}}]{Winter2018}%
  \BibitemOpen
  \bibfield  {author} {\bibinfo {author} {\bibfnamefont {S.~M.}\ \bibnamefont
  {Winter}}, \bibinfo {author} {\bibfnamefont {K.}~\bibnamefont {Riedl}},
  \bibinfo {author} {\bibfnamefont {D.}~\bibnamefont {Kaib}}, \bibinfo {author}
  {\bibfnamefont {R.}~\bibnamefont {Coldea}}, \ and\ \bibinfo {author}
  {\bibfnamefont {R.}~\bibnamefont {Valentí}},\ }\href {\doibase
  10.1103/PhysRevLett.120.077203} {\bibfield  {journal} {\bibinfo  {journal}
  {Phys. Rev. Lett.}\ }\textbf {\bibinfo {volume} {120}},\ \bibinfo {pages}
  {077203} (\bibinfo {year} {2018})}\BibitemShut {NoStop}%
\bibitem [{\citenamefont {Eichstaedt}\ \emph {et~al.}(2019)\citenamefont
  {Eichstaedt}, \citenamefont {Zhang}, \citenamefont {Laurell}, \citenamefont
  {Okamoto}, \citenamefont {Eguiluz},\ and\ \citenamefont
  {Berlijn}}]{Eichstaedt2019}%
  \BibitemOpen
  \bibfield  {author} {\bibinfo {author} {\bibfnamefont {C.}~\bibnamefont
  {Eichstaedt}}, \bibinfo {author} {\bibfnamefont {Y.}~\bibnamefont {Zhang}},
  \bibinfo {author} {\bibfnamefont {P.}~\bibnamefont {Laurell}}, \bibinfo
  {author} {\bibfnamefont {S.}~\bibnamefont {Okamoto}}, \bibinfo {author}
  {\bibfnamefont {A.~G.}\ \bibnamefont {Eguiluz}}, \ and\ \bibinfo {author}
  {\bibfnamefont {T.}~\bibnamefont {Berlijn}},\ }\href {\doibase
  10.1103/PhysRevB.100.075110} {\bibfield  {journal} {\bibinfo  {journal}
  {Phys. Rev. B}\ }\textbf {\bibinfo {volume} {100}},\ \bibinfo {pages}
  {075110} (\bibinfo {year} {2019})}\BibitemShut {NoStop}%
\bibitem [{\citenamefont {Laurell}\ and\ \citenamefont
  {Okamoto}(2020)}]{Laurell2020}%
  \BibitemOpen
  \bibfield  {author} {\bibinfo {author} {\bibfnamefont {P.}~\bibnamefont
  {Laurell}}\ and\ \bibinfo {author} {\bibfnamefont {S.}~\bibnamefont
  {Okamoto}},\ }\href {\doibase 10.1038/s41535-019-0203-y} {\bibfield
  {journal} {\bibinfo  {journal} {npj Quantum Materials}\ }\textbf {\bibinfo
  {volume} {5}} (\bibinfo {year} {2020}),\
  10.1038/s41535-019-0203-y}\BibitemShut {NoStop}%
\bibitem [{\citenamefont {Maksimov}\ and\ \citenamefont
  {Chernyshev}(2020)}]{SashaC}%
  \BibitemOpen
  \bibfield  {author} {\bibinfo {author} {\bibfnamefont {P.~A.}\ \bibnamefont
  {Maksimov}}\ and\ \bibinfo {author} {\bibfnamefont {A.~L.}\ \bibnamefont
  {Chernyshev}},\ }\href {\doibase 10.1103/PhysRevResearch.2.033011} {\bibfield
   {journal} {\bibinfo  {journal} {Phys. Rev. Research}\ }\textbf {\bibinfo
  {volume} {2}},\ \bibinfo {pages} {033011} (\bibinfo {year}
  {2020})}\BibitemShut {NoStop}%
\bibitem [{\citenamefont {Janssen}\ \emph {et~al.}(2017)\citenamefont
  {Janssen}, \citenamefont {Andrade},\ and\ \citenamefont
  {Vojta}}]{Janssen-Vojta2017}%
  \BibitemOpen
  \bibfield  {author} {\bibinfo {author} {\bibfnamefont {L.}~\bibnamefont
  {Janssen}}, \bibinfo {author} {\bibfnamefont {E.~C.}\ \bibnamefont
  {Andrade}}, \ and\ \bibinfo {author} {\bibfnamefont {M.}~\bibnamefont
  {Vojta}},\ }\href {\doibase 10.1103/PhysRevB.96.064430} {\bibfield  {journal}
  {\bibinfo  {journal} {Phys. Rev. B}\ }\textbf {\bibinfo {volume} {96}},\
  \bibinfo {pages} {064430} (\bibinfo {year} {2017})}\BibitemShut {NoStop}%
\bibitem [{\citenamefont {Suzuki}\ and\ \citenamefont
  {Suga}(2018)}]{Suzuki_models2018}%
  \BibitemOpen
  \bibfield  {author} {\bibinfo {author} {\bibfnamefont {T.}~\bibnamefont
  {Suzuki}}\ and\ \bibinfo {author} {\bibfnamefont {S.-i.}\ \bibnamefont
  {Suga}},\ }\href {\doibase 10.1103/PhysRevB.97.134424} {\bibfield  {journal}
  {\bibinfo  {journal} {Phys. Rev. B}\ }\textbf {\bibinfo {volume} {97}},\
  \bibinfo {pages} {134424} (\bibinfo {year} {2018})}\BibitemShut {NoStop}%
\bibitem [{\citenamefont {Kubota}\ \emph {et~al.}(2015)\citenamefont {Kubota},
  \citenamefont {Tanaka}, \citenamefont {Ono}, \citenamefont {Narumi},\ and\
  \citenamefont {Kindo}}]{KubotaPhysRevB2015}%
  \BibitemOpen
  \bibfield  {author} {\bibinfo {author} {\bibfnamefont {Y.}~\bibnamefont
  {Kubota}}, \bibinfo {author} {\bibfnamefont {H.}~\bibnamefont {Tanaka}},
  \bibinfo {author} {\bibfnamefont {T.}~\bibnamefont {Ono}}, \bibinfo {author}
  {\bibfnamefont {Y.}~\bibnamefont {Narumi}}, \ and\ \bibinfo {author}
  {\bibfnamefont {K.}~\bibnamefont {Kindo}},\ }\href {\doibase
  10.1103/PhysRevB.91.094422} {\bibfield  {journal} {\bibinfo  {journal} {Phys.
  Rev. B}\ }\textbf {\bibinfo {volume} {91}},\ \bibinfo {pages} {094422}
  (\bibinfo {year} {2015})}\BibitemShut {NoStop}%
\bibitem [{\citenamefont {Sears}\ \emph {et~al.}(2020)\citenamefont {Sears},
  \citenamefont {Chern}, \citenamefont {Kim}, \citenamefont {Bereciartua},
  \citenamefont {Francoual}, \citenamefont {Kim},\ and\ \citenamefont
  {Kim}}]{Sears_fm-Kitaev_exp2020}%
  \BibitemOpen
  \bibfield  {author} {\bibinfo {author} {\bibfnamefont {J.~A.}\ \bibnamefont
  {Sears}}, \bibinfo {author} {\bibfnamefont {L.~E.}\ \bibnamefont {Chern}},
  \bibinfo {author} {\bibfnamefont {S.}~\bibnamefont {Kim}}, \bibinfo {author}
  {\bibfnamefont {P.~J.}\ \bibnamefont {Bereciartua}}, \bibinfo {author}
  {\bibfnamefont {S.}~\bibnamefont {Francoual}}, \bibinfo {author}
  {\bibfnamefont {Y.~B.}\ \bibnamefont {Kim}}, \ and\ \bibinfo {author}
  {\bibfnamefont {Y.-J.}\ \bibnamefont {Kim}},\ }\href {\doibase
  10.1038/s41567-020-0874-0} {\bibfield  {journal} {\bibinfo  {journal} {Nature
  Physics}\ }\textbf {\bibinfo {volume} {16}},\ \bibinfo {pages} {837}
  (\bibinfo {year} {2020})}\BibitemShut {NoStop}%
\bibitem [{\citenamefont {Katsura}\ \emph {et~al.}(2010)\citenamefont
  {Katsura}, \citenamefont {Nagaosa},\ and\ \citenamefont {Lee}}]{MagnonHEth1}%
  \BibitemOpen
  \bibfield  {author} {\bibinfo {author} {\bibfnamefont {H.}~\bibnamefont
  {Katsura}}, \bibinfo {author} {\bibfnamefont {N.}~\bibnamefont {Nagaosa}}, \
  and\ \bibinfo {author} {\bibfnamefont {P.~A.}\ \bibnamefont {Lee}},\ }\href
  {\doibase 10.1103/PhysRevLett.104.066403} {\bibfield  {journal} {\bibinfo
  {journal} {Phys. Rev. Lett.}\ }\textbf {\bibinfo {volume} {104}},\ \bibinfo
  {pages} {066403} (\bibinfo {year} {2010})}\BibitemShut {NoStop}%
\bibitem [{\citenamefont {Matsumoto}\ and\ \citenamefont
  {Murakami}(2011{\natexlab{a}})}]{MagnonHEth2}%
  \BibitemOpen
  \bibfield  {author} {\bibinfo {author} {\bibfnamefont {R.}~\bibnamefont
  {Matsumoto}}\ and\ \bibinfo {author} {\bibfnamefont {S.}~\bibnamefont
  {Murakami}},\ }\href {\doibase 10.1103/PhysRevLett.106.197202} {\bibfield
  {journal} {\bibinfo  {journal} {Phys. Rev. Lett.}\ }\textbf {\bibinfo
  {volume} {106}},\ \bibinfo {pages} {197202} (\bibinfo {year}
  {2011}{\natexlab{a}})}\BibitemShut {NoStop}%
\bibitem [{\citenamefont {Matsumoto}\ and\ \citenamefont
  {Murakami}(2011{\natexlab{b}})}]{MagnonHEth3}%
  \BibitemOpen
  \bibfield  {author} {\bibinfo {author} {\bibfnamefont {R.}~\bibnamefont
  {Matsumoto}}\ and\ \bibinfo {author} {\bibfnamefont {S.}~\bibnamefont
  {Murakami}},\ }\href {\doibase 10.1103/PhysRevB.84.184406} {\bibfield
  {journal} {\bibinfo  {journal} {Phys. Rev. B}\ }\textbf {\bibinfo {volume}
  {84}},\ \bibinfo {pages} {184406} (\bibinfo {year}
  {2011}{\natexlab{b}})}\BibitemShut {NoStop}%
\bibitem [{\citenamefont {Shindou}\ \emph {et~al.}(2013)\citenamefont
  {Shindou}, \citenamefont {Ohe}, \citenamefont {Matsumoto}, \citenamefont
  {Murakami},\ and\ \citenamefont {Saitoh}}]{MagnonHEth5}%
  \BibitemOpen
  \bibfield  {author} {\bibinfo {author} {\bibfnamefont {R.}~\bibnamefont
  {Shindou}}, \bibinfo {author} {\bibfnamefont {J.-i.}\ \bibnamefont {Ohe}},
  \bibinfo {author} {\bibfnamefont {R.}~\bibnamefont {Matsumoto}}, \bibinfo
  {author} {\bibfnamefont {S.}~\bibnamefont {Murakami}}, \ and\ \bibinfo
  {author} {\bibfnamefont {E.}~\bibnamefont {Saitoh}},\ }\href {\doibase
  10.1103/PhysRevB.87.174402} {\bibfield  {journal} {\bibinfo  {journal} {Phys.
  Rev. B}\ }\textbf {\bibinfo {volume} {87}},\ \bibinfo {pages} {174402}
  (\bibinfo {year} {2013})}\BibitemShut {NoStop}%
\bibitem [{\citenamefont {Murakami}\ and\ \citenamefont
  {Okamoto}(2016)}]{Murakami2016}%
  \BibitemOpen
  \bibfield  {author} {\bibinfo {author} {\bibfnamefont {S.}~\bibnamefont
  {Murakami}}\ and\ \bibinfo {author} {\bibfnamefont {A.}~\bibnamefont
  {Okamoto}},\ }\href {\doibase 10.7566/JPSJ.86.011010} {\bibfield  {journal}
  {\bibinfo  {journal} {J. Phys. Soc. Jpn.}\ }\textbf {\bibinfo {volume}
  {86}},\ \bibinfo {pages} {011010} (\bibinfo {year} {2016})}\BibitemShut
  {NoStop}%
\bibitem [{\citenamefont {Fukui}\ \emph {et~al.}(2005)\citenamefont {Fukui},
  \citenamefont {Hatsugai},\ and\ \citenamefont {Suzuki}}]{bcapprox1}%
  \BibitemOpen
  \bibfield  {author} {\bibinfo {author} {\bibfnamefont {T.}~\bibnamefont
  {Fukui}}, \bibinfo {author} {\bibfnamefont {Y.}~\bibnamefont {Hatsugai}}, \
  and\ \bibinfo {author} {\bibfnamefont {H.}~\bibnamefont {Suzuki}},\
  }\href@noop {} {\bibfield  {journal} {\bibinfo  {journal} {Journal of the
  Physical Society of Japan}\ }\textbf {\bibinfo {volume} {74}},\ \bibinfo
  {pages} {1674} (\bibinfo {year} {2005})}\BibitemShut {NoStop}%
\bibitem [{\citenamefont {Park}\ and\ \citenamefont {Yang}(2019)}]{bcapprox2}%
  \BibitemOpen
  \bibfield  {author} {\bibinfo {author} {\bibfnamefont {S.}~\bibnamefont
  {Park}}\ and\ \bibinfo {author} {\bibfnamefont {B.-J.}\ \bibnamefont
  {Yang}},\ }\href@noop {} {\bibfield  {journal} {\bibinfo  {journal} {Physical
  Review B}\ }\textbf {\bibinfo {volume} {99}},\ \bibinfo {pages} {174435}
  (\bibinfo {year} {2019})}\BibitemShut {NoStop}%
\bibitem [{\citenamefont {Chern}\ \emph {et~al.}(2020)\citenamefont {Chern},
  \citenamefont {Kaneko}, \citenamefont {Lee},\ and\ \citenamefont
  {Kim}}]{Chern-YBK-semiclassics2020}%
  \BibitemOpen
  \bibfield  {author} {\bibinfo {author} {\bibfnamefont {L.~E.}\ \bibnamefont
  {Chern}}, \bibinfo {author} {\bibfnamefont {R.}~\bibnamefont {Kaneko}},
  \bibinfo {author} {\bibfnamefont {H.-Y.}\ \bibnamefont {Lee}}, \ and\
  \bibinfo {author} {\bibfnamefont {Y.~B.}\ \bibnamefont {Kim}},\ }\href
  {\doibase 10.1103/PhysRevResearch.2.013014} {\bibfield  {journal} {\bibinfo
  {journal} {Phys. Rev. Res.}\ }\textbf {\bibinfo {volume} {2}},\ \bibinfo
  {pages} {013014} (\bibinfo {year} {2020})}\BibitemShut {NoStop}%
\bibitem [{\citenamefont {Kittel}(2004)}]{Kittel}%
  \BibitemOpen
  \bibfield  {author} {\bibinfo {author} {\bibfnamefont {C.}~\bibnamefont
  {Kittel}},\ }\href
  {http://www.amazon.com/Introduction-Solid-Physics-Charles-Kittel/dp/047141526X/ref=dp_ob_title_bk}
  {\emph {\bibinfo {title} {Introduction to Solid State Physics}}},\ \bibinfo
  {edition} {8th}\ ed.\ (\bibinfo  {publisher} {Wiley},\ \bibinfo {year}
  {2004})\BibitemShut {NoStop}%
\bibitem [{\citenamefont {Liu}\ \emph {et~al.}(2021)\citenamefont {Liu},
  \citenamefont {Granados~del \'Aguila}, \citenamefont {Bhowmick},
  \citenamefont {Gan}, \citenamefont {Thu Ha~Do}, \citenamefont {Prosnikov},
  \citenamefont {Sedmidubsk\'y}, \citenamefont {Sofer}, \citenamefont
  {Christianen}, \citenamefont {Sengupta},\ and\ \citenamefont
  {Xiong}}]{LiePhysRevLett2021}%
  \BibitemOpen
  \bibfield  {author} {\bibinfo {author} {\bibfnamefont {S.}~\bibnamefont
  {Liu}}, \bibinfo {author} {\bibfnamefont {A.}~\bibnamefont {Granados~del
  \'Aguila}}, \bibinfo {author} {\bibfnamefont {D.}~\bibnamefont {Bhowmick}},
  \bibinfo {author} {\bibfnamefont {C.~K.}\ \bibnamefont {Gan}}, \bibinfo
  {author} {\bibfnamefont {T.}~\bibnamefont {Thu Ha~Do}}, \bibinfo {author}
  {\bibfnamefont {M.~A.}\ \bibnamefont {Prosnikov}}, \bibinfo {author}
  {\bibfnamefont {D.}~\bibnamefont {Sedmidubsk\'y}}, \bibinfo {author}
  {\bibfnamefont {Z.}~\bibnamefont {Sofer}}, \bibinfo {author} {\bibfnamefont
  {P.~C.~M.}\ \bibnamefont {Christianen}}, \bibinfo {author} {\bibfnamefont
  {P.}~\bibnamefont {Sengupta}}, \ and\ \bibinfo {author} {\bibfnamefont
  {Q.}~\bibnamefont {Xiong}},\ }\href {\doibase 10.1103/PhysRevLett.127.097401}
  {\bibfield  {journal} {\bibinfo  {journal} {Phys. Rev. Lett.}\ }\textbf
  {\bibinfo {volume} {127}},\ \bibinfo {pages} {097401} (\bibinfo {year}
  {2021})}\BibitemShut {NoStop}%
\bibitem [{\citenamefont {Lebert}\ \emph {et~al.}(2022)\citenamefont {Lebert},
  \citenamefont {Kim}, \citenamefont {Prishchenko}, \citenamefont {Tsirlin},
  \citenamefont {Said}, \citenamefont {Alatas},\ and\ \citenamefont
  {Kim}}]{LebertPhysRevB2022}%
  \BibitemOpen
  \bibfield  {author} {\bibinfo {author} {\bibfnamefont {B.~W.}\ \bibnamefont
  {Lebert}}, \bibinfo {author} {\bibfnamefont {S.}~\bibnamefont {Kim}},
  \bibinfo {author} {\bibfnamefont {D.~A.}\ \bibnamefont {Prishchenko}},
  \bibinfo {author} {\bibfnamefont {A.~A.}\ \bibnamefont {Tsirlin}}, \bibinfo
  {author} {\bibfnamefont {A.~H.}\ \bibnamefont {Said}}, \bibinfo {author}
  {\bibfnamefont {A.}~\bibnamefont {Alatas}}, \ and\ \bibinfo {author}
  {\bibfnamefont {Y.-J.}\ \bibnamefont {Kim}},\ }\href {\doibase
  10.1103/PhysRevB.106.L041102} {\bibfield  {journal} {\bibinfo  {journal}
  {Phys. Rev. B}\ }\textbf {\bibinfo {volume} {106}},\ \bibinfo {pages}
  {L041102} (\bibinfo {year} {2022})}\BibitemShut {NoStop}%
\bibitem [{\citenamefont {Li}\ \emph {et~al.}(2021)\citenamefont {Li},
  \citenamefont {Zhang}, \citenamefont {Said}, \citenamefont {Fabbris},
  \citenamefont {Mazzone}, \citenamefont {Yan}, \citenamefont {Mandrus},
  \citenamefont {Hal{\'{a}}sz}, \citenamefont {Okamoto}, \citenamefont
  {Murakami}, \citenamefont {Dean}, \citenamefont {Lee},\ and\ \citenamefont
  {Miao}}]{LiNatComm2021}%
  \BibitemOpen
  \bibfield  {author} {\bibinfo {author} {\bibfnamefont {H.}~\bibnamefont
  {Li}}, \bibinfo {author} {\bibfnamefont {T.~T.}\ \bibnamefont {Zhang}},
  \bibinfo {author} {\bibfnamefont {A.}~\bibnamefont {Said}}, \bibinfo {author}
  {\bibfnamefont {G.}~\bibnamefont {Fabbris}}, \bibinfo {author} {\bibfnamefont
  {D.~G.}\ \bibnamefont {Mazzone}}, \bibinfo {author} {\bibfnamefont {J.~Q.}\
  \bibnamefont {Yan}}, \bibinfo {author} {\bibfnamefont {D.}~\bibnamefont
  {Mandrus}}, \bibinfo {author} {\bibfnamefont {G.~B.}\ \bibnamefont
  {Hal{\'{a}}sz}}, \bibinfo {author} {\bibfnamefont {S.}~\bibnamefont
  {Okamoto}}, \bibinfo {author} {\bibfnamefont {S.}~\bibnamefont {Murakami}},
  \bibinfo {author} {\bibfnamefont {M.~P.~M.}\ \bibnamefont {Dean}}, \bibinfo
  {author} {\bibfnamefont {H.~N.}\ \bibnamefont {Lee}}, \ and\ \bibinfo
  {author} {\bibfnamefont {H.}~\bibnamefont {Miao}},\ }\href {\doibase
  10.1038/s41467-021-23826-1} {\bibfield  {journal} {\bibinfo  {journal}
  {Nature Communications}\ }\textbf {\bibinfo {volume} {12}} (\bibinfo {year}
  {2021}),\ 10.1038/s41467-021-23826-1}\BibitemShut {NoStop}%
\bibitem [{\citenamefont {Nagaosa}\ \emph {et~al.}(2010)\citenamefont
  {Nagaosa}, \citenamefont {Sinova}, \citenamefont {Onoda}, \citenamefont
  {MacDonald},\ and\ \citenamefont {Ong}}]{Nagaosa-RMP}%
  \BibitemOpen
  \bibfield  {author} {\bibinfo {author} {\bibfnamefont {N.}~\bibnamefont
  {Nagaosa}}, \bibinfo {author} {\bibfnamefont {J.}~\bibnamefont {Sinova}},
  \bibinfo {author} {\bibfnamefont {S.}~\bibnamefont {Onoda}}, \bibinfo
  {author} {\bibfnamefont {A.~H.}\ \bibnamefont {MacDonald}}, \ and\ \bibinfo
  {author} {\bibfnamefont {N.~P.}\ \bibnamefont {Ong}},\ }\href {\doibase
  10.1103/RevModPhys.82.1539} {\bibfield  {journal} {\bibinfo  {journal} {Rev.
  Mod. Phys.}\ }\textbf {\bibinfo {volume} {82}},\ \bibinfo {pages} {1539}
  (\bibinfo {year} {2010})}\BibitemShut {NoStop}%
\bibitem [{\citenamefont {Flebus}\ and\ \citenamefont
  {MacDonald}(2022)}]{Flebus-MacDonald2022}%
  \BibitemOpen
  \bibfield  {author} {\bibinfo {author} {\bibfnamefont {B.}~\bibnamefont
  {Flebus}}\ and\ \bibinfo {author} {\bibfnamefont {A.~H.}\ \bibnamefont
  {MacDonald}},\ }\href {\doibase 10.1103/PhysRevB.105.L220301} {\bibfield
  {journal} {\bibinfo  {journal} {Phys. Rev. B}\ }\textbf {\bibinfo {volume}
  {105}},\ \bibinfo {pages} {L220301} (\bibinfo {year} {2022})}\BibitemShut
  {NoStop}%
\bibitem [{\citenamefont {Chen}\ \emph {et~al.}(2020)\citenamefont {Chen},
  \citenamefont {Kivelson},\ and\ \citenamefont {Sun}}]{Chen-Kivelson2020}%
  \BibitemOpen
  \bibfield  {author} {\bibinfo {author} {\bibfnamefont {J.-Y.}\ \bibnamefont
  {Chen}}, \bibinfo {author} {\bibfnamefont {S.~A.}\ \bibnamefont {Kivelson}},
  \ and\ \bibinfo {author} {\bibfnamefont {X.-Q.}\ \bibnamefont {Sun}},\ }\href
  {\doibase 10.1103/PhysRevLett.124.167601} {\bibfield  {journal} {\bibinfo
  {journal} {Phys. Rev. Lett.}\ }\textbf {\bibinfo {volume} {124}},\ \bibinfo
  {pages} {167601} (\bibinfo {year} {2020})}\BibitemShut {NoStop}%
\bibitem [{\citenamefont {Guo}\ \emph {et~al.}(2022)\citenamefont {Guo},
  \citenamefont {Joshi},\ and\ \citenamefont {Sachdev}}]{Guo-Sachdev2022}%
  \BibitemOpen
  \bibfield  {author} {\bibinfo {author} {\bibfnamefont {H.}~\bibnamefont
  {Guo}}, \bibinfo {author} {\bibfnamefont {D.~G.}\ \bibnamefont {Joshi}}, \
  and\ \bibinfo {author} {\bibfnamefont {S.}~\bibnamefont {Sachdev}},\ }\href
  {\doibase 10.1073/pnas.2215141119} {\bibfield  {journal} {\bibinfo  {journal}
  {Proc. Natl. Acad. Sci. U.S.A.}\ }\textbf {\bibinfo {volume} {119}},\
  \bibinfo {pages} {e2215141119} (\bibinfo {year} {2022})}\BibitemShut
  {NoStop}%
\bibitem [{\citenamefont {Mandal}(2021)}]{mandal2021}%
  \BibitemOpen
  \bibfield  {author} {\bibinfo {author} {\bibfnamefont {I.}~\bibnamefont
  {Mandal}},\ }\href {\doibase 10.12693/APhysPolA.140.372} {\bibfield
  {journal} {\bibinfo  {journal} {Acta Phys. Pol. A}\ }\textbf {\bibinfo
  {volume} {140}},\ \bibinfo {pages} {372} (\bibinfo {year}
  {2021})}\BibitemShut {NoStop}%
\bibitem [{\citenamefont {Winter}()}]{Winter-comm}%
  \BibitemOpen
  \bibfield  {author} {\bibinfo {author} {\bibfnamefont {S.~M.}\ \bibnamefont
  {Winter}},\ }\href@noop {} {}\bibinfo {note} {Private
  communication}\BibitemShut {NoStop}%
\end{thebibliography}%


\clearpage
\setcounter{equation}{0}
\setcounter{figure}{0}
\setcounter{table}{0}
\makeatletter
\renewcommand{\theequation}{S\arabic{equation}}
\renewcommand{\thefigure}{S\arabic{figure}}
\renewcommand{\bibnumfmt}[1]{[#1]}
\renewcommand{\citenumfont}[1]{#1}

\onecolumngrid

\begin{center}
	\large{\textbf{Supplementary Materials for ``\thistitle"}}
\end{center}


\section{Details of the linear spin-wave theory}

The magnetic sector of the Hamiltonian is given in Eq.~\eqref{eq.model} in the main text,
which we repeat here:
\begin{equation}\label{smeq.model}
\begin{split}
  H_{\text{m}}=&\sum_{\langle ij\rangle_1\in\alpha}[J_1\vec{S}_i\cdot\vec{S}_j+KS_i^{\alpha}S_j^{\alpha}+\Gamma(S_i^{\beta}S_j^{\gamma}+S_i^{\gamma}S_j^{\beta})\\
&+\Gamma'(S_i^{\alpha}S_j^{\beta}+S_i^{\beta}S_j^{\alpha}+S_i^{\gamma}S_j^{\alpha}+S_i^{\alpha}S_j^{\gamma})]\\
  &+\sum_{\langle ij\rangle_3}J_3\vec{S}_i\cdot\vec{S}_j-\sum_i \frac{g\mu_B}{\hbar}\vec{h}\cdot\vec{S}_i,
\end{split}
\end{equation}
It contains five free parameters $J_1,~K,~\Gamma,~\Gamma',~J_3$ in the spin-spin couplings, and also the external magnetic field $\vec{h}$.
We note that the components $\alpha=(x,y,z)$ correspond to the Kitaev (``cubic'') axes, rather than the crystallographic $(a,b,c)$ axes. The relation between the two coordinate systems is given by: 
\begin{equation}\label{smeq.unitary}
\begin{split}
\begin{pmatrix}
S^x\\
S^y\\
S^z
\end{pmatrix}
=
\begin{pmatrix}
\sqrt{\frac{1}{6}} & -\sqrt{\frac{1}{2}} & \sqrt{\frac{1}{3}}\\
\sqrt{\frac{1}{6}} & \sqrt{\frac{1}{2}} & \sqrt{\frac{1}{3}}\\
-\sqrt{\frac{2}{3}} & 0 & \sqrt{\frac{1}{3}}
\end{pmatrix}
\begin{pmatrix}
S^a\\
S^b\\
S^c
\end{pmatrix}
.
\end{split}    
\end{equation}

We note in passing that while additional terms in the spin Hamiltonian have been considered in the literature (see e.g. Ref.~\cite{mandal2021}), the above $J_1$-$K$-$\Gamma$-$\Gamma'$-$J_3$ parametrization appears to be widely adopted~\cite{Kim2016,Winter2016,Janssen-Vojta2017,Suzuki_models2018,Laurell2020,Chern2021PRL,SashaC}. 

\textbf{Ordered phases and reference states}. For each parameter set, we first determine the magnetically ordered ground state by minimizing the classical energy.
Because the spins in \rucl\  form a zigzag order at a low field and transition to a polarized phase at a high field, 
we make the ground state ansatz that the ground-state spin configurations are captured by the two-sublattice unit cell of the (canted) zigzag-type  ordering, with four Ru atoms per unit cell (labelled A, B, C, D in Fig.~\ref{fig.kgmodel} (a)).
There are regions in the phase diagram where the classical calculations have shown that magnetic orders with larger unit cells may exist~\cite{YBK,Chern-YBK-semiclassics2020},
 however for the purpose of a manageable computation, we   ignore such regions 
 (which turn out to be very small in our parameter scans) and limit our considerations to where the ground states are captured by the four-site ansatz. 

Under the application of an external magnetic field (taken to be along the $a$-axis as in the experiments), these four spins   cant along the field direction, which can be parametrized by the polar and azimuthal angles $(\theta,\phi)$ in the $(abc)$ crystallographic coordinates, as shown in Fig.~\ref{fig.kgmodel} in the main text, such that:
$\theta_A =\theta_B=\theta_1,~\phi_A=\phi_B=\phi_1,
    \theta_C =\theta_D=\theta_2,~\phi_C=\phi_D=\phi_2$. 
For fixed model parameters and magnetic field, the classical energy thus becomes a function of the four angles $(\theta_1,\phi_1,\theta_2,\phi_2)$. After the classical energy minimization, the reference phases that we observe, in the order of increasing magnetic field, are as follows:
\begin{itemize}[itemsep=0.5ex,partopsep=0ex,parsep=0.5ex]
    \item Zigzag phase \rom{1} (ZZ1): canted zigzag phase with spins in the $(ac)$-plane, such that $\phi_1=\phi_2=0$ and $\theta_1\neq\theta_2$;
    \item Zigzag phase \rom{2} (ZZ2): canted zigzag phase with spins in the $(ab)$-plane, such that $\theta_1=\theta_2=\frac{\pi}{2}$ and $\phi_1=-\phi_2$;
    \item Polarized phase (PL) with spins along the $a$ field direction: $\phi_1=\phi_2=0$ and $\theta_1=\theta_2=\pi/2$.
\end{itemize}
These spin configurations are shown in Figs.~\ref{fig.kgmodel} (b,c,d) in the main text. In the small field region $0<h<h_{c1}$, the system is in the ZZ1 phase, in which the spins are situated in the $ac$ plane, see Fig.~\ref{fig.kgmodel}(b). With increasing field $h>h_{c1}$, the system transitions to the ZZ2 phase, where the spins lie in the $ab$ plane, see Fig.~\ref{fig.kgmodel}(c).
In the region $h_{c2}<h<h_{c3}$, the magnon band structure becomes unphysical (having imaginary eigenvalues), symptomatic of the failure of the zigzag ansatz to capture the true ground state, which may an indication of other four-spin order, or of the need to consider enlarged magnetic unit cell, such as for instance found in the semiclassical analysis~\cite{YBK,Chern-YBK-semiclassics2020}. Analyzing such enlarged unit cells is beyond the scope of the present work and we use the abbreviation UN in Fig.~\ref{fig.kgmodel} of the main text to represent the unknown phases. As mentioned in the previous paragraph, such unknown phases constitute only a very small region of the parameter regime we have surveyed in this work and we do not expect their existence to qualitatively alter our conclusions.
Finally, the system enters the  fully polarized phase for fields $h>h_{c3}$.

A side note is that we sometimes (depending on the model parameters) find a small region  of a noncollinear zigzag phase \rom{3} (ZZ3) where the spins are not confined to either the $ac$ nor $ab$ planes. Occasionally, we also observe a partially polarized (PPL) phase, in which all the spins are collinear but do not point along the magnetic field (this occurs due to strong spin-orbit coupling, for large $|\Gamma|$ or $|\Gamma'|$). 
Given the tiny regime   the ZZ3 and PPL phases, and the fact that unlike the previously discussed phases, they do not appear universally in all parameter sets, we focus in what follows on the three main phases ZZ1, ZZ2 and PL.

\vspace{2mm}
\textbf{Linear Spin Wave Theory.} In each of the  phases above, we perform the linear spin wave theory (LSWT) calculations to obtain the magnon band structure. The quantization axis (local $z$ direction) is chosen such that it coincides with the mean-field spin direction on a given site. In the following discussion, the tilde over the spin operators ($\tilde{S}_i$) indicates that these are in the local coordinate frame with the site-dependent quantization axis.

We then perform the standard Holstein-Primakoff transformation in this local   basis, expressing the spin operators in terms of the magnon creation and annihilation operators $a^\dagger$ and $a$:
\begin{equation}
\label{eq.hpa}
   \hat{\tilde{S}}_i^+= \sqrt{2S}\hat{a}_i^{\dagger},~\hat{\tilde{S}}_i^-= \sqrt{2S}\hat{a}_i^{\dagger},~\hat{\tilde{S}}_i^z=S-\hat{a}_i^{\dagger}\hat{a}_i.
\end{equation}
Upon the Fourier transformation to $k$-space, the Hamiltonian in Eq.~\eqref{smeq.model}  takes a quadratic form in the magnon operators:
\begin{equation}
H_{\text{m}}=\frac{1}{2}\sum_{\vec{k}}\,\Phi_{\vec{k}}^{\dagger}\,H(\vec{k})\,\Phi_{\vec{k}},
\end{equation}
where $H(\vec{k})$ is a $2N\times2N$ matrix.
Here $N$ is the number of sites in the magnetic unit cell. $N = 4$ for ZZ1 and ZZ2 orders, and $N=2$ for PL order. The matrix in the Nambu space is of the form
\begin{equation} 
H(\vec{k})
=
\begin{pmatrix}
H_{11}(\vec{k}) & H_{12}(\vec{k}) \\
H_{12}(-\vec{k})^* & H_{11}(-\vec{k})^{^\intercal} \\
\end{pmatrix}\ , 
\end{equation}
with $H_{11}(\vec{k})$ and $H_{12}(\vec{k})$ being $N\times N$ matrices. To be concrete, in the case of the PL phase, the matrix acts on the ket vector $\Psi_{\vec{k}}=(a_{\vec{k}},b_{\vec{k}},a_{-\vec{k}}^{\dagger},b_{-\vec{k}}^{\dagger})^{\intercal}$, and the block submatrices $H_{11}(\vec{k})$ and $H_{12}(\vec{k})$ are given by 
\begin{align}
H_{11}(\vec{k})
&=
\begin{pmatrix}
f_{aa}(\vec{k}) & f_{ab}(\vec{k}) \\
f_{ab}(\vec{k})^* & f_{aa}(\vec{k})\\
\end{pmatrix},\\
H_{12}(\vec{k})
&=
\begin{pmatrix}
0 & g_{ab}(\vec{k}) \\
g_{ab}(-\vec{k}) & 0\\
\end{pmatrix},
\end{align}   
\vspace{-2mm}
where the $\vec{k}$-dependent functions are
\begin{align}
    f_{aa}(\vec{k})&=\frac{1}{2} \left(2 \Gamma '+\Gamma -3 J_1-3 J_3-K\right)+h,\\
    f_{ab}(\vec{k})&=\frac{1}{4} e^{-i \left(-\frac{1}{2} \sqrt{3} k_a-\frac{k_b}{2}\right)} \left(\frac{\Gamma '}{3}+\frac{2 \Gamma }{3}+2 J_1+\frac{5
   K}{6}\right)
   +\frac{1}{4} e^{-i \left(\frac{\sqrt{3} k_a}{2}-\frac{k_b}{2}\right)} \left(\frac{\Gamma '}{3}+\frac{2 \Gamma }{3}+2 J_1+\frac{5
   K}{6}\right) \nonumber\\
   &+\frac{1}{4} e^{-i k_b} \left(\frac{4 \Gamma '}{3}-\frac{\Gamma }{3}+2 J_1+\frac{K}{3}\right) +\frac{1}{2} J_3 \left(e^{-i
   \left(k_b-\sqrt{3} k_a\right)}+e^{-i \left(\sqrt{3} k_a+k_b\right)}+e^{2 i k_b}\right),\\
   g_{ab}(\vec{k})&=\frac{1}{4} e^{-i k_b} \left(-\frac{4 \Gamma '}{3}-\frac{5 \Gamma }{3}-\frac{K}{3}\right) +e^{-i \left(-\frac{1}{2} \sqrt{3} k_a-\frac{k_b}{2}\right)} \left(\frac{1}{4} \left(-\frac{7 \Gamma '}{3}-\frac{2 \Gamma
   }{3}+\frac{K}{6}\right)-\frac{i \left(\Gamma '-\Gamma +K\right)}{2 \sqrt{6}}\right) \nonumber\\
   &+e^{-i \left(\frac{\sqrt{3} k_a}{2}-\frac{k_b}{2}\right)} \left(\frac{1}{4} \left(-\frac{7 \Gamma '}{3}-\frac{2 \Gamma
   }{3}+\frac{K}{6}\right)+\frac{i \left(\Gamma '-\Gamma +K\right)}{2 \sqrt{6}}\right).
\end{align}

\vspace{-2mm}
In the case of ZZ1 and ZZ2 phases,  $\Psi_{\vec{k}}=(a_{\vec{k}},b_{\vec{k}},c_{\vec{k}},d_{\vec{k}},a_{-\vec{k}}^{\dagger},b_{-\vec{k}}^{\dagger},c_{-\vec{k}}^{\dagger},d_{-\vec{k}}^{\dagger})^{\intercal}$ is the Nambu ket-vector composed of the magnon operators on sublattices A, B, C, D in Fig.~\ref{fig.kgmodel}(a). The band dispersions $\epsilon_{n\vec{k}}$ and eigenvectors $|n,\vec{k}\rangle$ at each $\vec{k}$ point are obtained by the similarity (Bogoliubov) transformation. 

\begin{figure}[!ht]
    \centering
    \includegraphics[width=0.4\textwidth]{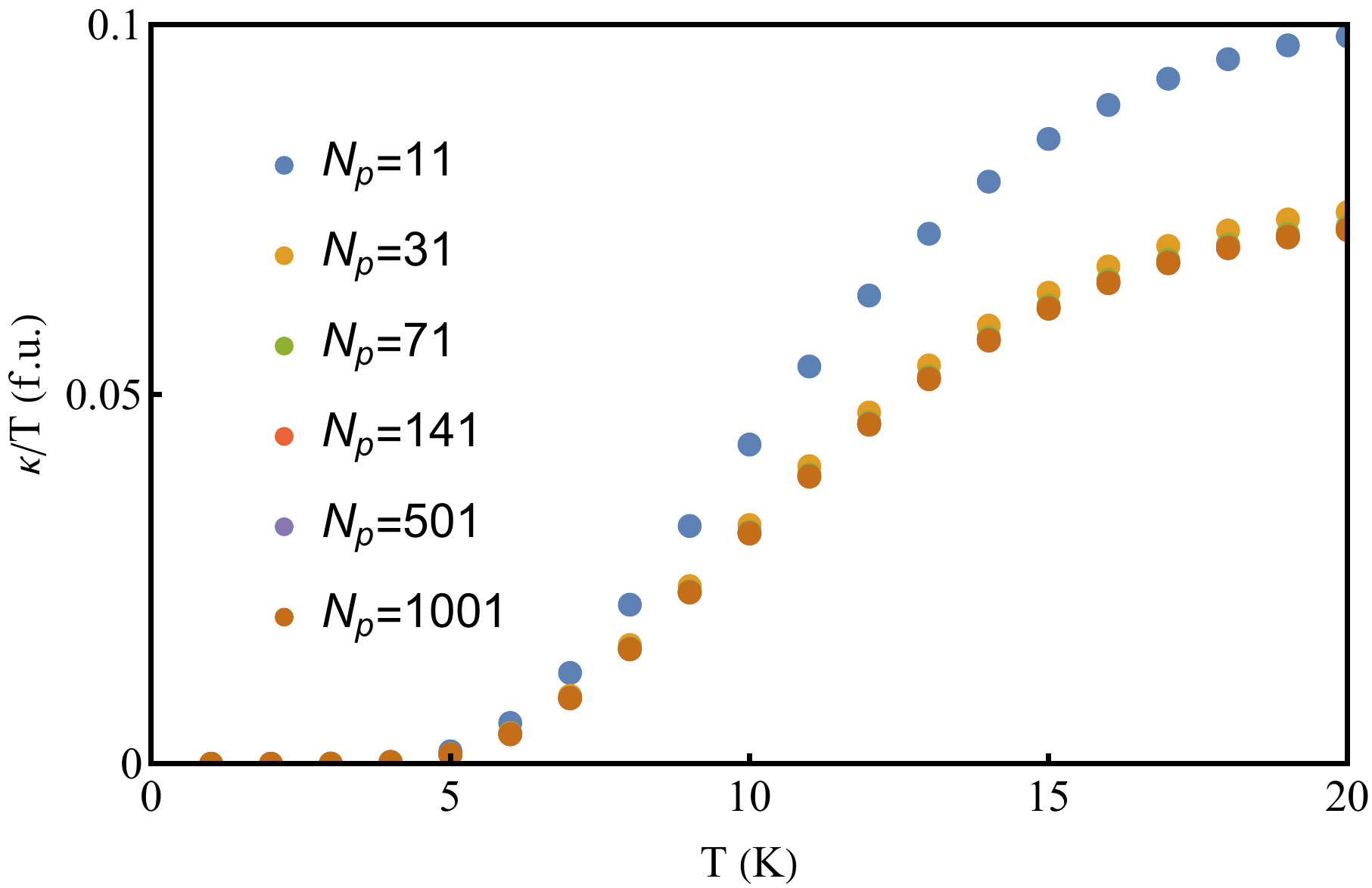}
    \caption{
    Demonstration of convergence of computed  $\kappa_{xy}/T$ for model parameters $(J_1,J_3,K,\Gamma,\Gamma')=(0, 0, -7.2, 2.2, -0.2)$~meV and field $h=10$~T. The k-space mesh $N_p\times N_p$ was chosen to perform numerical integration in  Eq.~\eqref{smeq.kxyfu}, with varying $N_p$ as shown in the legend. The results demonstrate that convergence is reached for $N_p\ge 71$.
    In the scan over the parameter space (Figs.~\ref{fig.pl},\ref{fig.zz}), we used $N_p =  71$.
    }
\label{fig.convm}
\end{figure}

In order to compute the thermal Hall conductivity, one must compute the integral  of the Berry curvature weighted by the $c_2(f(\epsilon_{n\vec{k}}))$ function, as explained in Eq.~\eqref{eq.kxy} in the main text. For numerical purposes, the integral  is replaced by a discrete sum (same as Eq.~3\eqref{eq.kxyfu} in the main text) as follows:
\begin{equation}
\label{smeq.kxyfu}
\left.\frac{\kappa_{xy}^{2D}}{T}\right\vert_{\text{f.u.}}=\frac{\kappa_{xy}^{3D}d}{T}\times\frac{6\hbar}{\pi k_B^2}=\frac{3}{2\pi^3}\!\sum_{n,\vec{k}\in BZ(2D)}\!\!c_2(f(\epsilon_{n\vec{k}_0}))\phi_{n\vec{k}},
\end{equation}
Here $\phi_{n\vec{k}}$ is the Berry flux through a small plaquette formed by $\vec{\delta}_1=(Q_a/N_p,0)$ and $\vec{\delta}_2=(0,Q_b/N_p)$ with $Q_a=2\pi/\sqrt{3}$ and $Q_b=4\pi/3$ (for $N=2$) or $2\pi/3$ (for $N=4$), respectively. The elementary flux is given by \cite{bcapprox1,bcapprox2}: 
\begin{equation}
\begin{split}
   \phi_{n\vec{k}}=&-Arg[\langle n,\vec{k}|\Sigma_{2N}|n,\vec{k}+\vec{\delta}_1\rangle\langle n,\vec{k}+\vec{\delta}_1|\Sigma_{2N}|n,\vec{k}+\vec{\delta}_1+\vec{\delta}_2\rangle
   \langle n,\vec{k}+\vec{\delta}_1+\vec{\delta}_2|\Sigma_{2N}|n,\vec{k}+\vec{\delta}_2\rangle\langle n,\vec{k}+\vec{\delta}_2|\Sigma_{2N}|n,\vec{k}\rangle], 
\end{split}
\end{equation}
where $\Sigma_{2N}=\sigma_3 \otimes I_{N\times N}$ with the Pauli matrix $\sigma_3=\mathrm{diag}\{1,-1\}$
%
 and $I_{N\times N}$  the identity matrix of size $N$. In our calculations of Eq.~\eqref{smeq.kxyfu}, we use the discrete mesh $N_p\times N_p$, and we show that $N_p=71$ 
 is sufficiently large for the discretized integration to converge numerically
 in the pure magnon case (see Fig.~\ref{fig.convm}).
  
\section{Comparison of the model parameters against prior theoretical models and experiments}
\label{sec.S-params}

To compare with the experimental data, we choose three sets of parameters where we find   the largest values of $|\kappa_{xy}/T|$ in the scanned parameter range. These parameters sets (ps) belong to the PL, ZZ1 and ZZ2 phases, respectively, in the in-plane field of $h=10$~T relevant for comparison with experiments:
\begin{itemize}
    \item 
    ps1:  $(K,J_1,J_3,\Gamma,\Gamma')=(-7.2,0,0.2,0.2,-0.4)$ meV,
    \item 
    ps2:  $(K,J_1,J_3,\Gamma,\Gamma')=(-7.2,0,0.8,0,-0.2)$ meV, 
    \item 
    ps3:  $(K,J_1,J_3,\Gamma,\Gamma')=(-7.2,-2,3,5,3.6)$ meV,
\end{itemize}
These parameter sets are labeled by the colored diamond symbols in Figs.~\ref{fig.pl}(c,d) and Figs.~\ref{fig.zz}(c,d). 

In addition, we compare our model parameters with those used by previous studies: our ps1 and ps2 lie close to the parameter set used by Zhang \textit{et al.}~\cite{YBK} (labeled by a cross in Figs.~\ref{fig.pl} and \ref{fig.zz}), while 
ps3 lies on the edge of what Maksimov and Chernyshev call the ``realistic parameter regime'' in their work~\cite{SashaC}, shown with a dashed rectangle in Figs.~\ref{fig.pl} and \ref{fig.zz}. In all cases, we find that the maximum contributions of magnons to $|\kappa^{2D}_{xy}|/T$ do not exceed about 0.3 fermionic units, significantly lower than the experimentally reported values of thermal Hall conductivity around 0.5 fermionic unit~\cite{Matsuda-quantized,ONG}.

\vspace{0mm}
\begin{center}
\textbf{Field and temperature dependence and comparison with experiments}
\end{center}
The temperature $T$ and magnetic field $h$ dependence of $\kappa_{xy}/T$ in these three parameter sets are plotted against the experimental data in Figs.~\ref{fig.pl} and \ref{fig.zz} in the main text. 
 In the case of the polarized phase ps1, the field dependence of $|\kappa_{xy}|/T$ is qualitatively different from the experimental data, monotonically decreasing with the increasing magnetic field. The possibility of being in the polarized phase at 10~T is thus ruled out, as explained in the main text. 
 
 For the remaining two parameters sets (ps2 and ps3), under the experimentally relevant conditions $2<T<6$~K and $6<h<7$~T, we find that the 
parameter set ps2 qualitatively matches the trends in the experimental data  of $\kappa_{xy}/T$. 
Near $T=10$ K and $h=10$ T, $\kappa_{xy}/T$ continues to increase for ps2 (as is the case experimentally); whereas for ps3 it starts to decline slowly. The parameter set ps2 (in the ZZ1 phase) thus captures the experimental behavior of $\kappa_{xy}/T$ qualitatively in the temperature and magnetic regions of the measurement. However, we find the strength of $\kappa_{xy}/T$ in all cases to be much smaller than the experimental result. This mismatch shows that the intrinsic magnon contribution alone cannot fully account for the 
experimentally measured thermal Hall conductivity in \rucl. This corroborates the experimental suggestions~\cite{taillefer-phonons2022} that other sources, in particular phonons, must contribute significantly to the thermal Hall effect in this material.

\section{Magnetoelastic coupling and intrinsic phonon contribution to $ \kappa_{xy}$}

Given the suggested importance of acoustic phonons in contributing to the thermal Hall in \rucl~\cite{taillefer-phonons2022}, in this section we describe the modeling of phonons and phonon-magnon coupling, in order to explore their intrinsic contributions to the thermal Hall conductivity. The extrinsic (scatterer-dependent) contribution will be discussed separately in section~\ref{sec.S-ext}.


\subsection{Phonon Model}
\begin{figure}[!ht]
    \centering
    \includegraphics[width=0.7\textwidth]{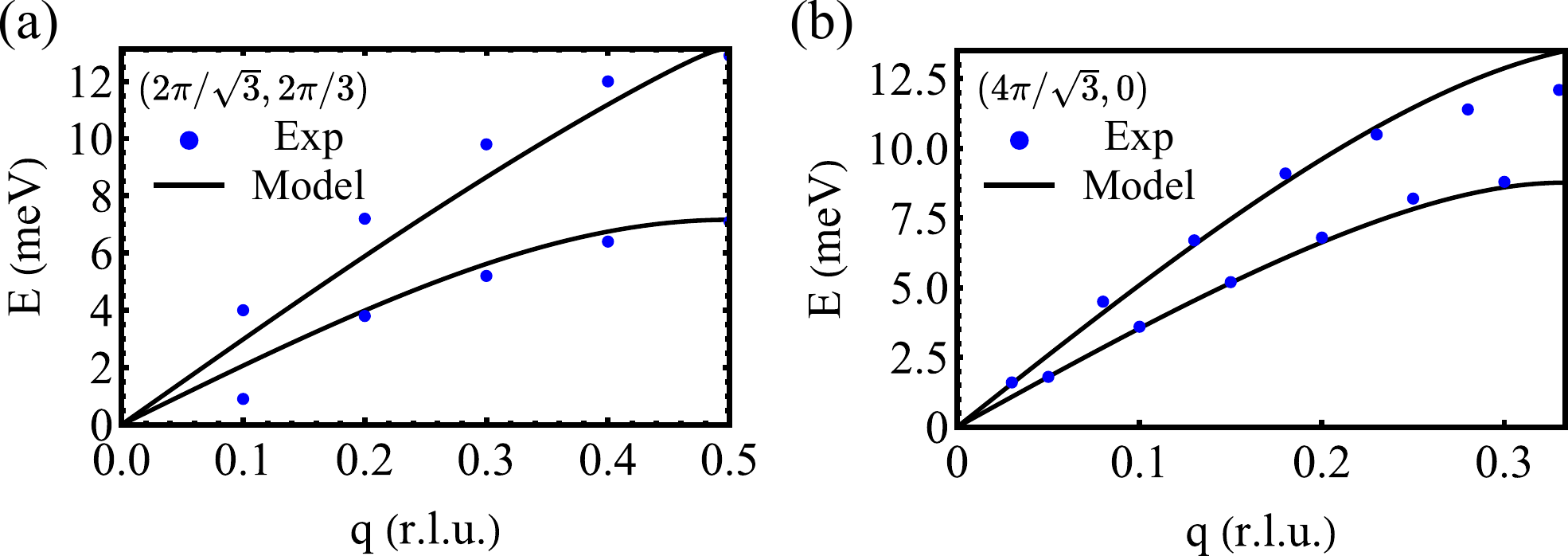}
    \caption{
    Acoustic phonon dispersion  in our model compared to experimental measurement in Ref.~\cite{LebertPhysRevB2022}. \vspace{-8mm}
    }
\label{fig.phononband}
\end{figure}

To describe the phonon modes in $\alpha$-RuCl$_3$, we use a nearest neighbor elastic model \cite{LiePhysRevLett2021} on the honeycomb lattice:


\begin{equation}   
    H_{\text{PH}}=\frac{1}{2M}\sum_i\vec{\Pi}_i^2+\frac{M}{2}\sum_i\vec{u}_i^{\intercal}\Phi_0\vec{u}_i+\frac{M}{2}\sum_{i\in A}\sum_{\alpha}\vec{u}_i\Phi_{\alpha}\vec{u}_{i+\alpha}+\frac{M}{2}\sum_{i\in B}\sum_{\alpha}\vec{u}_{i-\alpha}\Phi_{\alpha}\vec{u}_{i}, 
\end{equation}
where $\vec{u}_i=(u_i^a,u_i^b)$ and $\vec{\Pi}_i=(\Pi_i^a,\Pi_i^b)$ are   displacement and its momentum at site $i$, $\alpha=\vec{a}_1 (x),~\vec{a}_2 (y),~\vec{a}_3 (z)$ are vectors of three types of nearest-neighbor bonds. The matrix $\Phi_{\alpha}$ are given by:
\begin{equation}
\begin{aligned}
\Phi_{\vec{\alpha}_1}
=
\begin{pmatrix}
-\gamma_1+\frac{\sqrt{3}}{2}\gamma_2 & \frac{1}{2}\gamma_2 \\
\frac{1}{2}\gamma_2 & -\gamma_1-\frac{\sqrt{3}}{2}\gamma_2\\
\end{pmatrix},~
\Phi_{\vec{\alpha}_2}
=
\begin{pmatrix}
-\gamma_1-\frac{\sqrt{3}}{2}\gamma_2 & \frac{1}{2}\gamma_2 \\
\frac{1}{2}\gamma_2 & -\gamma_1+\frac{\sqrt{3}}{2}\gamma_2\\
\end{pmatrix},~
\Phi_{\vec{\alpha}_3}
=
\begin{pmatrix}
-\gamma_1 & -\gamma_2 \\
-\gamma_2 & -\gamma_1\\
\end{pmatrix},~
\Phi_0
=
\begin{pmatrix}
3\gamma_1 & 0 \\
0 & 3\gamma_1\\
\end{pmatrix}.
\end{aligned}     
\end{equation}
Here, $\gamma_{1,2}$ are free parameters allowed by the hexagonal symmetry of the lattice,  whose values we fix to match the experiments~\cite{LebertPhysRevB2022}.
After the second quantization and Fourier transform, the phonon Hamiltonian can be written in a quadratic form:
\begin{equation}
H_{\text{PH}}=\frac{1}{2}\sum_{\vec{k}}\Phi_{\vec{k}}^{\dagger}H_{\text{ph}}(\vec{k})\Phi_{\vec{k}},
\end{equation}
where $H_{\text{ph}}(\vec{k})$ is a $8\times8$ matrix (or $16\times16$ matrix for zigzag order unit cell) and $\Phi_{\vec{k}}=(A^a_{\vec{k}},A^b_{\vec{k}},...,A_{-\vec{k}}^{a\dagger},A_{-\vec{k}}^{b\dagger},...)$, $A_{\vec{k}}$ represents the phonon annihilation operator.
The matrix $H_{\text{PH}}(\vec{k})$ is given by
\begin{equation}
\begin{aligned}
H_{\text{PH}}(\vec{k})
=
\begin{pmatrix}
H_{\text{PH}11}(\vec{k}) & H_{\text{PH}12}(\vec{k}) \\
H_{\text{PH}12}(-\vec{k})^* & H_{\text{PH}11}(-\vec{k})^{^\intercal} \\
\end{pmatrix}\ ,
\end{aligned} 
\end{equation}
written in terms of the block matrices 
\begin{equation}
\begin{aligned}
H_{\text{PH}11}(\vec{k})
=
\begin{pmatrix}
E & 0 & f_{AxBx}(\vec{k},E,c) & f_{AxBy}(\vec{k},E,c) \\
0 & E & f_{AyBx}(\vec{k},E,c) & f_{AyBy}(\vec{k},E,c) \\
f_{AxBx}^*(\vec{k},E,c) & f_{AyBx}^*(\vec{k},E,c) & E & 0 \\
f_{AxBy}^*(\vec{k},E,c) & f_{AyBy}^*(\vec{k},E,c) & 0 & E \\
\end{pmatrix}\ ,
\end{aligned} 
\end{equation}
\begin{equation}
\begin{aligned}
H_{\text{PH}12}(\vec{k})
=
\begin{pmatrix}
0 & 0 & f_{AxBx}(\vec{k},E,c) & f_{AxBy}(\vec{k},E,c) \\
0 & 0 & f_{AyBx}(\vec{k},E,c) & f_{AyBy}(\vec{k},E,c) \\
f_{AxBx}(-\vec{k},E,c) & f_{AyBx}(-\vec{k},E,c) & 0 & 0 \\
f_{AxBy}(-\vec{k},E,c) & f_{AyBy}(-\vec{k},E,c) & 0 & 0 \\
\end{pmatrix} . 
\end{aligned} 
\end{equation}
Here the functions used are
\begin{equation}
   f_{AxBx}(\vec{k},E,c)=\frac{1}{2}\left[\left(-\frac{E}{3} + \frac{\sqrt{3}c}{2}\right)e^{-i(-\frac{\sqrt{3}c}{2}k_a - \frac{1}{2} k_b)}+ \left(-\frac{E}{3} - \frac{\sqrt{3}c}{2}\right)
    e^{-i(\frac{\sqrt{3}c}{2}k_a - \frac{1}{2} k_b)}  -\frac{E}{3} e^{-ik_b}\right], 
\end{equation}
\begin{equation}
   f_{AyBy}(\vec{k},E,c)=\frac{1}{2}\left[\left(-\frac{E}{3} - \frac{\sqrt{3}c}{2}\right)e^{-i(-\frac{\sqrt{3}c}{2}k_a - \frac{1}{2} k_b)}+ \left(-\frac{E}{3} + \frac{\sqrt{3}c}{2}\right)
    e^{-i(\frac{\sqrt{3}c}{2}k_a - \frac{1}{2} k_b)} -\frac{E}{3} e^{-ik_b}\right], 
\end{equation}
\begin{equation}
    f_{AxBy}(\vec{k},E,c)=f_{AyBx}(\vec{k},E,c)=\frac{1}{2}\left[\frac{c}{2}e^{-i(-\frac{\sqrt{3}c}{2}k_a - \frac{1}{2} k_b)}+ \frac{c}{2}
    e^{-i(\frac{\sqrt{3}c}{2}k_a - \frac{1}{2} k_b)} + (-c) e^{-ik_b}\right],
\end{equation}
where $E=\sqrt{3\gamma_1}\hbar$ and $c=\gamma_2\hbar/\sqrt{3\gamma_1}$. To capture the behavior of phonons measured in experiments by Lebert \textit{et al. }\cite{LebertPhysRevB2022}, we perform a linear square fit to the experimental phonon dispersions, and  obtain $E=13.5$ and $c=2.6$. The band structure is shown in Fig.~\ref{fig.phononband}. 

\vspace{-6mm}
\subsection{Magnon-phonon coupling}
We now introduce the coupling between phonons and magnons, which is essential to endow the phonons with chirality and hence enable their intrinsic contribution to the thermal Hall effect (for extrinsic contribution, see Section~\ref{sec.S-ext}).
A distance dependence of the superexchange interactions $J(R_i-R_j)$ between Ru$^{3+}$ ions leads naturally to the magnetoelastic coupling in a general form
\beq
\label{smeq.ME}
\begin{split}
H_\text{ME}& = \sum_{i,j} S_i^\alpha S_j^\beta\,  (\vec{u}_i - \vec{u}_j)\cdot\vec{{\nabla}}_{\mathbf{r}_{ij}}
\, J^{\alpha\beta}(\mathbf{r}_{ij})
=\sum_{i,j}\vec{S}_i^{\intercal}\frac{\partial {\bm{J}}(r_{ij})}{\partial r_{ij}}\vec{S}_j\, [\hat{\vec{r}}_{ij}\cdot(\vec{u}_j-\vec{u}_i)]\\
&=\sum_{i,j}\tilde{\vec{S}}_i^{\intercal}\bm{R}_i^{\intercal}\frac{\partial {\bm{J}}(r_{ij})}{\partial r_{ij}}\bm{R}_j\tilde{\vec{S}}_j\, [\hat{\vec{r}}_{ij}\cdot(\vec{u}_j-\vec{u}_i)].
\end{split}
\eeq
where $\vec{u}_i$ is the displacement of the $i^\text{th}$ ions from its equilibrium position, and $\hat{\vec{r}}_{ij}$ is the unit vector between site $i$ and $j$.
The bold $\bm{J}$ is the matrix notation of $J^{\alpha\beta}$.
In the second last line, the spin components $S^{\alpha}$ and the matrix $\bm{J}$ are expressed in the Kitaev coordinate system $(xyz)$.
In the last line, the spins $\tilde{\vec{S}}$ are in the 
local coordinate determined by the magnetic order, where $\tilde{S}^z$ is aligned with the direction of ordered spin.
Their relation is given by ${\vec{S}}_i=\bm{R}_i\tilde{\vec{S}}_i$, where
the rotational matrix $\bm{R}_i$ is
\begin{equation} 
\begin{split}
\bm{R}_i =
\begin{pmatrix}
\cos{\theta_i}\cos{\phi_i} & -\sin{\phi_i} & \sin{\theta_i}\cos{\phi_i}\\
\cos{\theta_i}\sin{\phi_i} & \cos{\phi_i} & \sin{\theta_i}\sin{\phi_i}\\
-\sin{\theta_i} & 0 & \cos{\theta_i}
\end{pmatrix},
\end{split}    
\end{equation}
where $(\theta_i,\phi_i)$ are the ordered spin polar angles in Kitaev coordinates.



The above equation~\eqref{smeq.ME} is sufficient to derive all the specific terms for the magnon-phonon coupling, which will contain many free parameters of the form $\frac{\partial \bm{J}(r_{ij})}{\partial r_{ij}}$.  
Here we take the term from $\tilde{S}_{A,i}^x\tilde{S}_{B,i+\vec{a}_1}^z$ as an example. Its magnon-phonon coupling term is:
\begin{equation}
    \left[\bm{R}_{A,i}^{\intercal}\frac{\partial \bm{J}(r_{\vec{a}_1})}{\partial r_{\vec{a}_1}}\bm{R}_{B,i+\vec{a}_1}\right]_{13} ~
    \times~\hat{\vec{a}}_1\cdot(\vec{u}_{B,i+\vec{a}_1}-\vec{u}_{A,i})  ~\times ~\tilde{S}_{A,i}^x\tilde{S}_{B,i+\vec{a}_1}^z,
\end{equation}
where the magneto-elastic coupling can be parametrized as follows:
\begin{equation}
\label{smeq.g-couplings}
   \sqrt{\frac{\hbar^2}{2M}}\frac{\partial \bm{J}(r_{\vec{a}_1})}{\partial r_{\vec{a}_1}}
=
\sqrt{\frac{\hbar^2}{2M}}\begin{pmatrix}
    \frac{\partial J_1}{\partial r}+\frac{\partial K}{\partial r} & \frac{\partial \Gamma'}{\partial r} & \frac{\partial \Gamma'}{\partial r}\\
    \frac{\partial \Gamma'}{\partial r} & \frac{\partial J_1}{\partial r} & \frac{\partial \Gamma}{\partial r}\\
    \frac{\partial \Gamma'}{\partial r} & \frac{\partial \Gamma}{\partial r} & \frac{\partial J_1}{\partial r}
    \end{pmatrix}
    \equiv \begin{pmatrix}
    g_1+g_2 & g_4 & g_4\\
    g_4 & g_1 & g_3\\
    g_4 & g_3 & g_1
    \end{pmatrix}.
\end{equation}
We then perform the standard Holstein-Primakoff transformation and write these displacements in terms of the phonon operators
$u_i^\gamma\sim (A_{i}^{\gamma\dagger} + A_{i}^{\gamma})$ (with polarization $\gamma$). 
This results in the hybridization between the magnons and phonons:
\begin{equation}
\begin{split}
    & \frac{1}{\sqrt{2E}}S^{\frac{3}{2}} \left[ \bm{R}_{A,i}^{\intercal}\sqrt{\frac{\hbar^2}{2M}}\frac{\partial \bm{J}(r_{\vec{a}_1})}{\partial r_{\vec{a}_1}}\bm{R}_{B,i+\vec{a}_1}  \right]_{13}
    \times  \\
    & \left[-\frac{\sqrt{3}}{2}(B_{i+\vec{a}_1}^{a\dagger}+B_{i+\vec{a}_1}^a-A_{i}^{a\dagger}-A_{i}^a)-\frac{1}{2}(B_{i+\vec{a}_1}^{b\dagger}+B_{i+\vec{a}_1}^b-A_{i}^{b\dagger}-A_{i}^b)\right](a_i+a_i^{\dagger}).
    \end{split}
\end{equation}
%
Other phonon-magnon coupling terms can be obtained in a similar fashion.
Now the total Hamiltonian is given by
\begin{equation}
\label{smeqn.total.ham}
    H=H_{\text{M}}+H_{\text{PH}}+H_{\text{ME}}. 
\end{equation}
In the case of zigzag phase, it takes a quadratic form in the magnon and phonon operators:
\begin{equation}
    H=\frac{1}{2}\sum_{\vec{k}}\Phi_{\vec{k}}^{\dagger}H(\vec{k})\Phi_{\vec{k}},
\end{equation}
where $H(\vec{k})$ is a $24\times24$ matrix and $\Psi_{\vec{k}}=(a_{\vec{k}},...,A^a_{\vec{k}},A^b_{\vec{k}},...,a_{-\vec{k}}^{\dagger},...,A_{-\vec{k}}^{a\dagger},A_{-\vec{k}}^{b\dagger},...)$, $a_{\vec{k}}$ and $A_{\vec{k}}$ representing the magnon and phonon annihilation operators. 
 
\vspace{-2mm}
\subsection{Parameter Fitting from Acoustic Phonon Softening}

\begin{figure}[!ht]
    \centering
\includegraphics[width=0.6\textwidth]{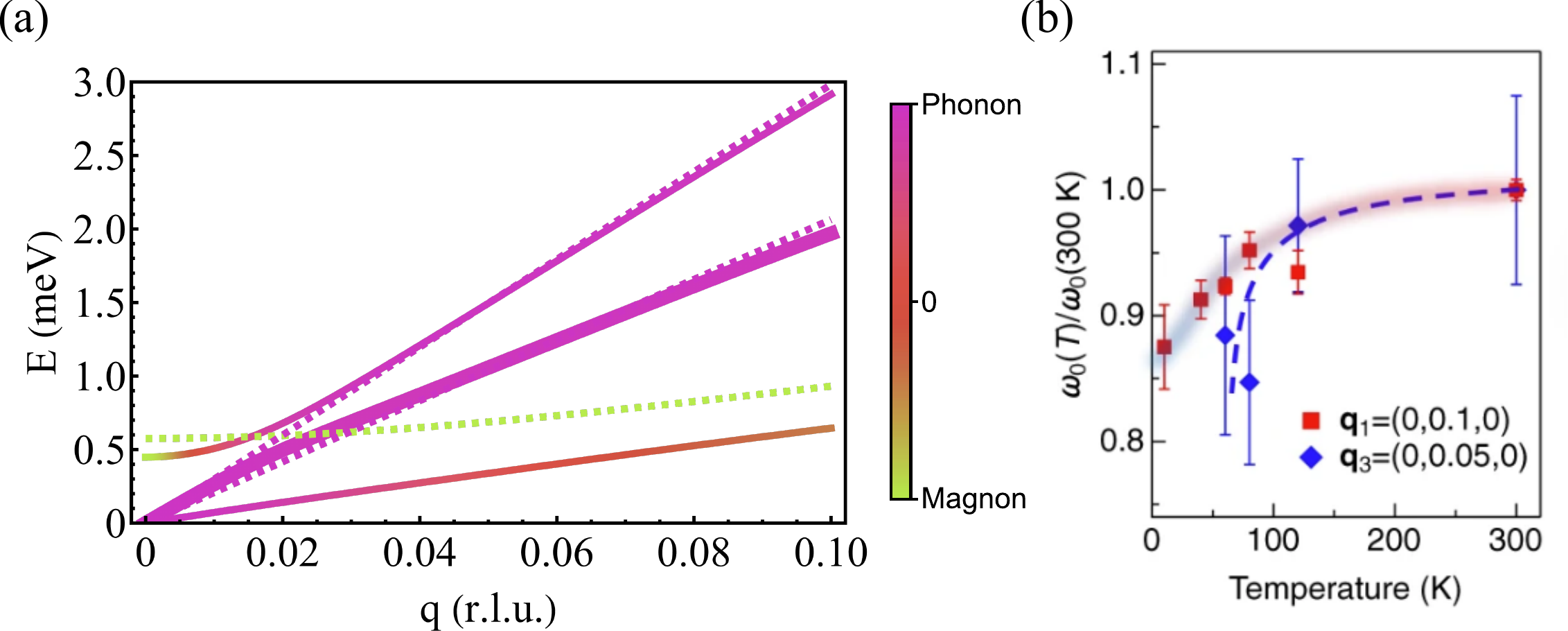}
    \caption{
    (a) Zoomed-in view of the magnon and phonon dispersion before and after including the magneto-elastic couplings, for parameters given in Eqs.~(\ref{eq.mpcoupling1},\ref{eq.mpcoupling2}).
    (b) Phonon softening measured in experiments in Ref.~\cite{LiNatComm2021}. This panel is taken from Fig.~\ref{fig.phonon}(e) of  Ref.~\cite{LiNatComm2021} directly.
    }
\label{fig.phonon.soft}
\end{figure}

The phonon-magnon coupling contains a few free parameters $g_1, g_2, g_3, g_4$ defined in Eq.~\eqref{smeq.g-couplings}. 
In order to fit these parameters, we refer to the recent experiment by Li \textit{et al} in Ref.~\cite{LiNatComm2021}, where the phonon dispersions are measured at both high temperatures (in the paramagnetic phase) and low temperatures (ordered phase) (see Fig.~\ref{fig.phonon.soft}(c)). 
The difference in these two scenarios is that the magnons are non-existent in the high-temperature paramagnetic phase, but interact with phonons at low temperatures.
Hence, the downward shift of the phonon dispersion (i.e. phonon softening) seen in the low-temperature data is
explicitly the result of the band anti-crossing, due to the phonon-magnon coupling, as shown in Fig.~\ref{fig.phonon.soft}(a). Hence, the magnitude of the phonon softening can be used to determine the magnitude of the magnon-phonon couplings, for parameter set ps2, given in Eqs.~(\ref{eq.mpcoupling1},\ref{eq.mpcoupling2}).
For the magnon sector, we use the parameter set 2 (ps2, see Section~\ref{sec.S-params}): 
\begin{equation}
\label{eq.mpcoupling1}
\{K,J_1,J_3,\Gamma,\Gamma'\}=\{-7.2, 0, 0.8, 0, -0.2 \},\qquad
h = 10 \text{~T}.
\end{equation}

While the experimental data is insufficient to obtain a unique fit, using a simplifying assumption that all $g$ coefficients are of the same order of magnitude.
In the main text, we use 
\begin{equation}
\label{smeq.mpcoupling2mt}
   g\equiv  g_1=g_2=-g_3=-g_4=4,
\end{equation}
to  qualitatively reproduce the phonon softening seen in the experiment. Please note that our units (defined in Eq.~\eqref{smeq.g-couplings}) are such that $g=1$ corresponds to $\partial \bm{J}(r_{\vec{a}_1})/\partial r_{\vec{a}_1}=4.9$~\text{meV/\AA}. We recognize that in reality the coefficients $g_i$ will likely have different magnitude, as shown by Winter \emph{et al.}~\cite{Winter-comm}, however the above simplifying assumption is sufficient to illustrate the effect that the magneto-elastic coupling has on the thermal Hall effect.

We also find that 
the maximal $g_i$'s without inducing a phase transition are
\begin{equation}
\label{eq.mpcoupling2}
   g\equiv   g_1=g_2=-g_3=-g_4=6.
\end{equation}

\begin{figure}[ht!]
    \centering
    \includegraphics[width=0.6\textwidth]{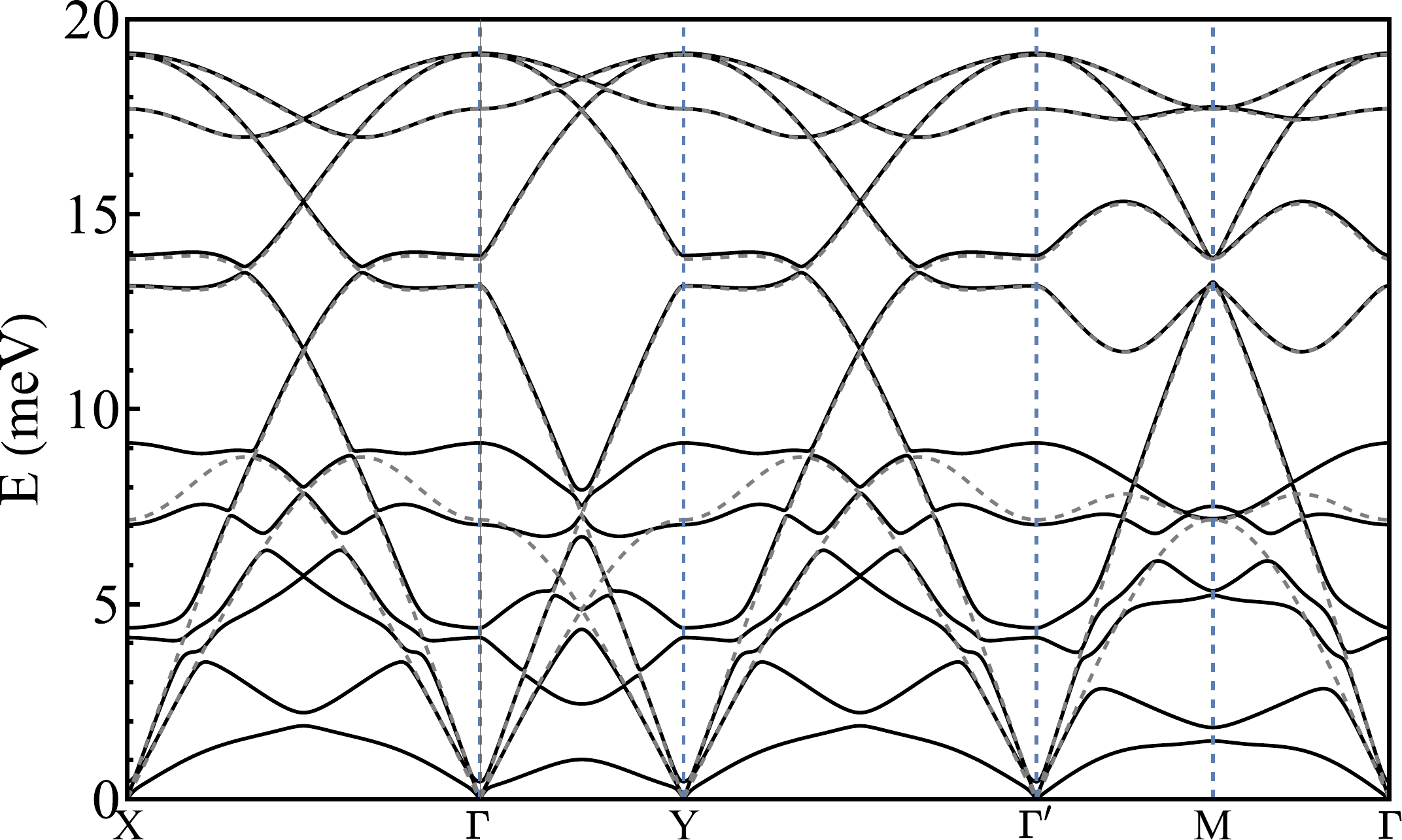}
    \caption{
    Band dispersion of coupled phonon and magnon modes along high-symmetry lines of the Brillouin zone.
    The computation is done for the parameter set 2 (ps2) given in Eqs.~(\ref{eq.mpcoupling1},\ref{eq.mpcoupling2}).
    Dashed lines are phonon dispersion without magneto-elastic couplings, and the solid lines are dispersion after including them.   
    }
\label{fig.magnon-phonon}
\end{figure}

Having thus determined all the coupling parameters in the Hamiltonian (Eq.~\ref{smeqn.total.ham}), we compute the dispersions of the coupled magnon and phonon branches, which are shown in
Fig.~\ref{fig.magnon-phonon} for $g_i$'s taking values in Eq.~\eqref{eq.mpcoupling2}. 
Figs.~\ref{fig.phonon.soft}(a) shows the zoomed-in view of the phonon and magnon dispersions before (dashed lines) and after (solid lines) turning on the magneto-elastic coupling, which is to be compared with the experimental data in Figs.~\ref{fig.phonon.soft}(b).

\begin{figure}[!ht]
    \centering
    \includegraphics[width=0.4\textwidth]{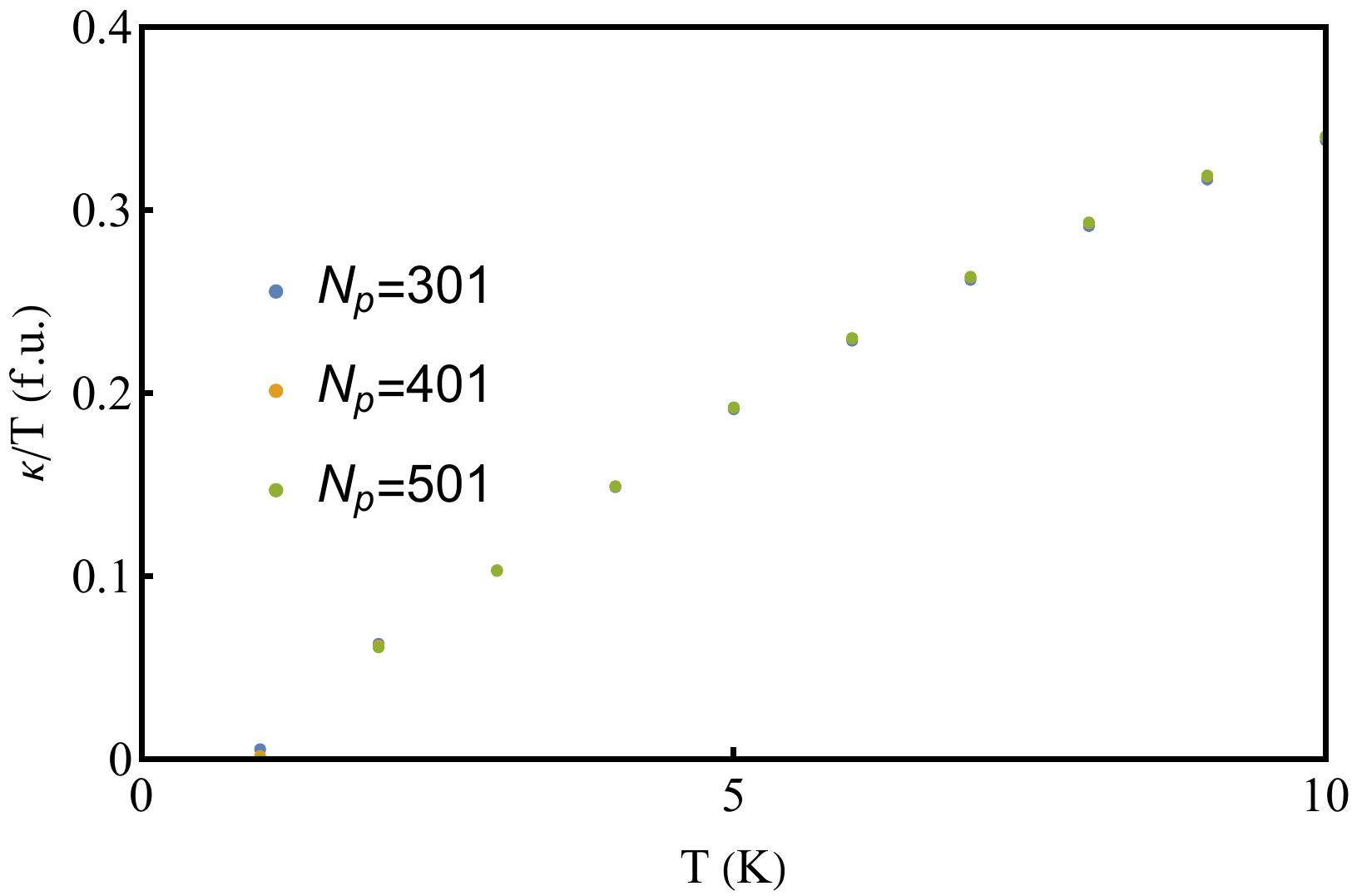}
    \caption{
    Demonstration of convergence of computed  $\kappa_{xy}/T$ (for parameter set ps2 and $g=6$, Eqs.~(\ref{eq.mpcoupling1},\ref{eq.mpcoupling2})). The k-space mesh $N_p\times N_p$ was chosen to perform numerical integration in  Eq.~\eqref{smeq.kxyfu}, with varying $N_p$. The results demonstrate that convergence is reached for $N_p\ge 401$. 
    In the calculation of Fig.~\ref{fig.phonon} in the main text, we used $N_p =  401$.
    }
\label{fig.convmp}
\end{figure}

After having obtained the hybridized magnon and phonon bands, we then compute the $\kappa_{xy}$ again, which now includes intrinsic phonon contributions.
For $g_i$'s taking values in Eq.~\eqref{smeq.mpcoupling2mt}, the result is shown in Fig.~\ref{fig.phonon} in the main text. 
For $g_i$'s taking the largest limiting values in Eq.~\eqref{eq.mpcoupling2}, the result is shown in Fig.~\ref{fig.phononappdix} below. 
The convergence of our calculation for $\kappa_{xy}$, which involves numerical integration over the Brillouin zone in Eq.~\eqref{smeq.kxyfu}, is shown in Fig.~\ref{fig.convmp},  for the case of Eq.~\eqref{eq.mpcoupling2}.

\vspace{-2mm}
\subsection{Limits on the strength of magneto-elastic coupling}\label{sec.g-limits}

Our final note of this section is that the coupling strength between magnons and phonons cannot be arbitrarily large.
A coupling too strong will either drive the system out of the presumed magnetic order and/or induce a lattice distortion. 
In terms of the original band theory,
this is manifested in unphysical band dispersions for values of magneto-elastic constants $g$ in Eq.~\eqref{smeq.g-couplings} that are too large. 

Given the complexity of the 12-band Hamiltonian (4 magnons and 8 phonon branches in the magnetic unit cell), we illustrate this point in the simplified picture.
Let us consider one phonon and one  magnon branch at a particular momentum.
Both the phonon and the magnon 
can be thought of as harmonic oscillators. 
In the classical picture, they are scalar variables $U$ and $S$ sitting in two quadratic potential wells 
\begin{equation}
    H = \omega_U U^2 + \omega_S S^2
    = (U, S) \begin{pmatrix}
        \omega_U & 0 \\ 0 & \omega_S
    \end{pmatrix}\begin{pmatrix}
        U\\ S
    \end{pmatrix},
\end{equation} 
where $\omega_U$ and $\omega_S$ correspond to the frequencies of the harmonic oscillators.
As long as $\omega_U$ and $\omega_S$  are positive, the ground state for the system is a stable minimum $U=S = 0$.

Upon introduction of the interaction between the two harmonic oscillators
\begin{equation}
    H' = \omega_U U^2 + \omega_S S^2 + 2g US
    = (U, S) \begin{pmatrix}
        \omega_U & g \\ g & \omega_S
    \end{pmatrix}\begin{pmatrix}
        U\\ S
    \end{pmatrix}.
\end{equation} 
The two eigenvalues of the resulting matrix are given by
\begin{equation}
\omega_\pm = \frac{1}{2}\left(
\omega_S+ \omega_U \pm \sqrt{(\omega_S-\omega_U)^2 + 4g^2}
\right).
\end{equation}
We can see here that when $ g^2 >   \omega_S \omega_U$, the lowest eigenvalue $\omega_-$ will become negative, indicating that the corresponding quantum eigenstate (a linear combination of the $U$ and $S$ modes) can have an arbitrarily large occupation number, resulting in the unbounded negative total energy (unless one considers higher-order terms in the Hamiltonian on physical grounds). 
This indicates that the hybridized magnon-phonon eigenmode will condense, drive the system into a different phase.
In our calculation, such types of phase transitions are not considered, nor are they realized experimentally in \rucl~\cite{LiNatComm2021}.

\section{Other forms of intrinsic phonon thermal Hall effect}


In addition to magnon-phonon coupling resulting in the Berry curvature for phonons, there may be other, intrinsic contributions of phonons to the thermal Hall effect. Fundamentally, any such mechanism requires phonons to obtain a chiral component, thus resulting in non-reciprocity of the thermal transport coefficients in an applied magnetic field. It has been proposed~\cite{Flebus-MacDonald2022} that one such mechanism may originate from the Lorentz force on ions (in our case, Ru$^{3+}$ cations and Cl$^{-}$ anions), coupling their longitudinal in-phase motion with an out-of-phase transverse motion, and thus endowing acoustic phonons with a chiral component. This is often termed ``intrinsic skew-scattering,'' specific to the crystal itself rather than the extrinsic properties of the phonon scatterers. 

Such intrinsic contribution would  however be strongly temperature dependent -- this is because the acquired chiral component of the acoustic phonons is strongly $k$-dependent, vanishing as $k^2$. Since the phonon momentum $k$ is inversely proportional to its de Broglie thermal wavelength, the resulting chiral component vanishes like $T^2$ at low temperatures. 
Using similar arguments, the authors of Refs.~\cite{Chen-Kivelson2020} and \cite{Guo-Sachdev2022} came to the conclusion that $\kappa_{xy}\propto T^4$. However, the experimentally measured ratio $\kappa_{xy}/\kappa_{xx}$ in \rucl is very weakly temperature dependent~\cite{taillefer-phonons2022}, thus indicating that such instrinsic skew-scattering mechanism is negligible, if at all present, in \rucl. 
That is why we turn our attention to \textit{extrinsic} mechanisms of phonon thermal Hall effect to explain the relatively large (and not quantized) experimentally measured value of $\kappa_{xy}/T$ in \rucl~\cite{taillefer-phonons2022,ONG}.

\section{Extrinsic contributions to the Hall effect}\label{sec.S-ext}

\subsection{Untenability of extrinsic skew scattering in \rucl}

One source of extrinsic phonon Hall effect is due to phonons scattering off of magnetic impurities, described phenomenologically by the \textit{extrinsic} skew scattering time $\tau_\text{skew}$ that depends on the concentration of magnetic impurities/defects. The corresponding contribution to the thermal Hall conductivity can be written as~\cite{Nagaosa-RMP} 
\beq
\kappa_{xy}^\text{skew} \sim 
\frac{1}{3} C_v v^2 \tau_\text{ph} (\tau_\text{ph} \tau_\text{skew}^{-1}),
\eeq
where $C_v$ is the specific heat, $v$ is the acoustic phonon velocity and $\tau_\text{ph}$ denotes the phonon scattering mean-free time. Written in terms of the same phenomenological parameters, the longitudinal thermal conductivity due to phonons is given by the well known formula~\cite{Kittel} 
\beq
\kappa_{xx} \sim \frac{1}{3} C_v v^2 \tau_\text{ph}.
\eeq
Taking the ratio of these two equations, we conclude  that 
 the thermal Hall angle $\kappa_{xy}/\kappa_{xx}\propto \tau_\text{ph}/\tau_{\text{skew}}$. Generally, one expects $\tau_\text{skew} \ll \tau_\text{ph}$ accounting for the smallness of the Hall angle. More importantly, the concentration of magnetic impurities or vacancies being a very much sample-dependent quantity, one expects the Hall angle to vary greatly from sample to sample, contradicting the observation that the Hall angle falls in the range $0.0003-0.0010$ for samples grown under different conditions in Ref.~\cite{taillefer-phonons2022}. The same logic was used by Guo \textit{et al.}~\cite{Guo-Sachdev2022} to argue that impurity skew scattering cannot explain the largely sample-independent thermal Hall effect observed in the hole-doped cuprate superconductor La$_{2-x}$Sr$_x$CuO$_4$. 

\vspace{-4mm}
\subsection{Side-jump scattering}

The other well known mechanism for electronic Hall effect ($\sigma_{xy}$) in metals is the so-called \textit{side-jump scattering}~\cite{Nagaosa-RMP}.
Guo \textit{et al.} have 
recently developed a formalism to describe the analogous effect for thermal Hall effect in insulating magnets~\cite{Guo-Sachdev2022}. While their motivation was primarily the large thermal Hall angle observed in the cuprates, the mechanism applies equally to \rucl. 
Without unnecessary duplication, we refer the reader to the original study in Ref.~\cite{Guo-Sachdev2022}, quoting here the main conclusions relevant to the present work.

It is reasonable to assume that the individual defects, off which phonons scatter, are coupled magnetically to the nearby Ru$^{3+}$ spins $\mathbf{S}$ as follows 
\beq
H_\text{def} = \sum_{R \text{ near } R_0} J(R-R_0)\, \mathbf{S}(R)\cdot\boldsymbol{\sigma}_\text{def},
\eeq
where summation is over ionic positions $R$ neighboring the defect. Following Guo \textit{et al.}, we represent the defect by a two-level system  $H_\text{def}=\Delta\sigma^z$, with the effective splitting
\beq
\Delta = \sum_{R \text{ near } R_0} J(R-R_0)\, \mathbf{S}(R).
\eeq
In zero field, this splitting is approximately zero (assuming the spherical symmetry of the defect with a sufficiently large radius), however in an applied magnetic field $h$, in the canted zigzag phase (ZZ1 or ZZ2), we expect $\Delta \propto h$. 

Because of the strong spin-orbit coupling in \rucl, the phonon momentum that is proportional to 
 the elastic strain $\eps^{ij}\sim \partial_i u^j$   couples to the defect spin as follows:
\beq
H_\text{ph-def} = K_{ij,\alpha} \eps^{ij}\sigma^\alpha.
\eeq
Following the derivation in Ref.~\cite{Guo-Sachdev2022}, we arrive at the formula for the extrinsic (E) thermal Hall conductivity
\beq
\kappa_{xy}^{E}/T \sim \frac{\Delta^4}{\tau_{ph}^{-1} T^3} \Phi\left(\frac{\Delta}{T}\right) \frac{\bar{K}^2}{v^3},
\eeq
where $v$ is the phonon velocity and $\bar{K}$ is an appropriate skew-symmetric combination of the matrix elements of $K_{ij,\alpha}$ (see Ref.~\cite{Guo-Sachdev2022}) for more detail. The universal function $\Phi(x)$ of the ratio $x=\Delta/T$ captures the distribution of two-level splittings on defects: if all defects have identical $\Delta$, then $\Phi(x) = 1/\sinh(x)$, and if the splittings are drawn from a distribution, then the function must be averaged over this distribution (which generally results in a power-law in $T$). 

Crucially, it follows that this extrinsic Hall conductivity is proportional to the phonon mean-free path $\ell=v \tau_\text{ph}$, as is the longitudinal thermal conductivity $\kappa_{xx}$, thus explaining the apparent sample-independent Hall angle $\theta_H^E=\kappa_{xy}^{E}/\kappa_{xx}$ in the experiments~\cite{taillefer-phonons2022}.

We proceed to phenomenologically determine the extrinsic Hall angle $\theta_H^E$ from the experimental data in Ref.~\cite{taillefer-phonons2022} at high temperatures $T\gtrsim |K|/k_B \approx 80K$, above the magnon bandwidth where the effects of the Kitaev physics and associated Berry curvature are unimportant:
\beq
\left.\frac{\kappa_{xy}}{\kappa_{xx}}\right|_\text{high $T$} =\theta_H^E.
\eeq
\noindent
We arrive at $\theta_H^E \approx (0.6 \pm 0.2) \times10^{-3}$.

By contrast, at low temperatures of the order $T\sim 10-15$~K where the interpretation of the thermal Hall measurements on \rucl is disputed, both the intrinsic (due to the Berry curvature) and extrinsic contributions to $\kappa_{xy}$ must be taken into account. In this low-temperature region
\beq
\left.\frac{\kappa_{xy}}{\kappa_{xx}}\right|_\text{low $T$} = \frac{\kappa_{xy}^I + \kappa_{xy}^E}{\kappa_{xx}} = \theta_H^E \left( 1 + \eta^{-1}\right),
\eeq
where $\eta\equiv \kappa_{xy}^E/\kappa_{xy}^I$ is the phenomenological ratio of the extrinsic and intrinsic   contributions.
From the analysis of the data in Ref.~\cite{taillefer-phonons2022}, we thus obtain $\eta=1.2\pm0.5$, with the uncertainty related to the spread of the experimental data among the (five) samples. This range of obtained $\eta$ ratios is used to determine the shaded blue region in Fig.~\ref{fig.phonon}(b) in the main text.

\begin{figure}[!ht]
    \centering
    \includegraphics[width=0.37\textwidth]{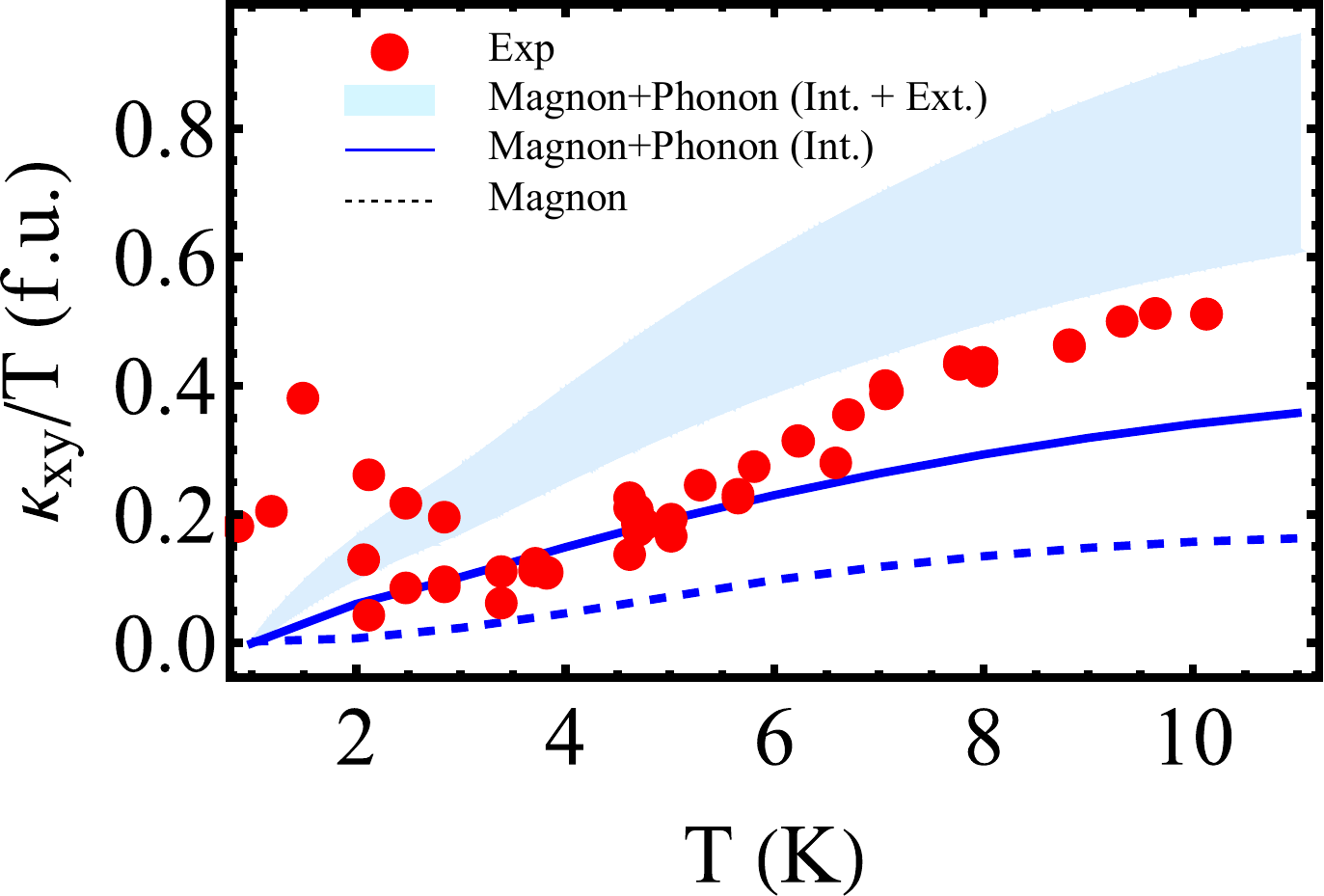}
    \caption{ 
    Total computed $\kappa_{xy}/T$,  contributed from different sources for parameters given in Eqs.~(\ref{eq.mpcoupling1},\ref{eq.mpcoupling2}), compared with the experimental data (red circles) from Ref.~\cite{ONG}. Here, a larger value of the ME coupling $g=6$ is used than in the main text. 
    The magnon and phonon intrinsic component $\kappa^{I}_{xy}$ (solid line), summed together with the extrinsic phonon contribution $\kappa^{E}_{xy}$, is indicated with the blue shaded region (whose width is given by the experimental uncertainty in determining $\kappa^{E}_{xy}=\eta \kappa^{I}_{xy}$, see text).
    For these parameters, the theoretical thermal Hall effect is stronger than measured in the experiment, as a proof of principle that the model is capable of reproducing large $\kappa_{xy}/T$, provided extrinsic contributions are taken into account.} 
    \label{fig.phononappdix}
\end{figure}

\vspace{3mm}
In the main text, a value of the magneto-elastic (ME) coupling g=4 (defined in Eqs.~\eqref{smeq.mpcoupling2mt} and \eqref{smeq.g-couplings}) was used as it provided a reasonable match to the experimental data. Here, we would like to remark that if instead one used the largest allowed value of $g\approx 6$ (see section~\ref{sec.g-limits} for the definition), the intrinsic magnon+phonon contribution becomes even larger, resulting in $\kappa^I_{xy}$ as large as 0.36 fermionic units at  $T=10$~K, as shown with a solid blue line in Fig.~\ref{fig.phononappdix}. This is still below the experimental value (red circles in Fig.~\ref{fig.phononappdix}), indicating the necessity to include extrinsic phonon contribution as discussed above. What this demonstrates however is that, upon including said contributions, the resulting $\kappa^{2D}_{xy}(T)$ would fall inside the shaded blue region in Fig.~\ref{fig.phononappdix}, exceeding the experimental values for the given set of model parameters (ps2, see section~\ref{sec.S-params}).
Hence, even if the model parameters are such that they do not exactly coincide with ps2 chosen to maximize the intrinsic $\kappa^I_{xy}$,  
a significant region in the parameter space may host strong enough thermal Hall effect that will match with the experiment.


\end{document}